\documentclass[floatfix,reprint,amsmath,amssymb,aps,superscriptaddress]{revtex4-1}
\usepackage[mathscr]{euscript}
 \let\mathscr\relax
\usepackage[scr]{rsfso}

\usepackage{float}
\usepackage{graphicx}
\usepackage{xcolor}
\usepackage{xparse}
\usepackage[caption=false]{subfig}
\usepackage{dcolumn}
\usepackage{bm}
\usepackage{amsmath}
\usepackage{physics}
\usepackage{array}
\usepackage{bbold}
\usepackage{commath}
\usepackage{braket}
\usepackage{dsfont}
\usepackage{tikz}
\usepackage{hyperref}
\hypersetup{colorlinks=true, citecolor=blue, urlcolor=blue, linkcolor=blue}

\ExplSyntaxOn

\keys_define:nn { miguel/label }
 {
  label   .tl_set:N = \l_miguel_label_tl,
  unknown .code:n   = \clist_put_right:Nx \l_miguel_label_clist
                       { \l_keys_key_tl = \exp_not:n { #1 } }
 }
\clist_new:N \l_miguel_label_clist
\box_new:N \l_miguel_label_box
\box_new:N \l_miguel_label_image_box

\NewDocumentCommand{\xincludegraphics}{O{}m}
 {
  \tl_clear:N \l_miguel_label_tl
  \clist_clear:N \l_miguel_label_clist
  \keys_set:nn { miguel/label } { #1 }
  \tl_if_empty:NTF \l_miguel_label_tl
   {
    \miguel_includegraphics:Vn \l_miguel_label_clist { #2 }
   }
   {
    \hbox_set:Nn \l_miguel_label_image_box
     {
      \miguel_includegraphics:Vn \l_miguel_label_clist { #2 }
     }
    \hbox_set:Nn \l_miguel_label_box
     {
      \skip_horizontal:n { 3pt }
      \fcolorbox{white}{white}{\footnotesize \tl_use:N \l_miguel_label_tl}
     }
    \leavevmode
    \box_use:N \l_miguel_label_image_box
    \skip_horizontal:n { -\box_wd:N \l_miguel_label_image_box }
    \hbox_overlap_right:n
     {
      \box_move_up:nn
       {
        \box_ht:N \l_miguel_label_image_box - 
        \box_ht:N \l_miguel_label_box - 3pt
       }
       { \box_use_drop:N \l_miguel_label_box }
     }
    \skip_horizontal:n { \box_wd:N \l_miguel_label_image_box }
   }
 }

\cs_new_protected:Nn \miguel_includegraphics:nn
 {
  \includegraphics[#1]{#2}
 }
\cs_generate_variant:Nn \miguel_includegraphics:nn { V }

\ExplSyntaxOff

\begin{document}

\title{Improving Gaussian channel simulation using non-unity gain heralded quantum teleportation}

\author{Biveen~Shajilal}
\thanks{These two authors contributed equally}
\affiliation{A*STAR Quantum Innovation Centre (Q.InC), Institute of Materials Research and Engineering (IMRE), Agency for Science, Technology and Research (A*STAR), 2 Fusionopolis Way, Singapore, 138634, Republic of Singapore}
\affiliation{Centre for Quantum Computation and Communication Technology, Research School of Engineering,
The Australian National University, Canberra, ACT 2601, Australia}
\affiliation{Centre for Quantum Computation and Communication Technology, Department of Quantum Science and Technology, The Australian National University, Canberra ACT 2601, Australia}

\author{Lorc{\'a}n~O.~Conlon}
\thanks{These two authors contributed equally}
\affiliation{A*STAR Quantum Innovation Centre (Q.InC), Institute of Materials Research and Engineering (IMRE), Agency for Science, Technology and Research (A*STAR), 2 Fusionopolis Way, Singapore, 138634, Republic of Singapore}
\affiliation{Centre for Quantum Computation and Communication Technology, Department of Quantum Science and Technology, The Australian National University, Canberra ACT 2601, Australia}

\author{Angus Walsh}
\affiliation{Centre for Quantum Computation and Communication Technology, Department of Quantum Science and Technology, The Australian National University, Canberra ACT 2601, Australia}

\author{Spyros~Tserkis}
\affiliation{Photonic Inc., Coquitlam, British Columbia, Canada}

\author{Jie~Zhao}
\affiliation{Centre for Quantum Computation and Communication Technology, Department of Quantum Science and Technology, The Australian National University, Canberra ACT 2601, Australia}

\author{Jiri~Janousek}
\affiliation{Centre for Quantum Computation and Communication Technology, Research School of Engineering,
The Australian National University, Canberra, ACT 2601, Australia}
\affiliation{Centre for Quantum Computation and Communication Technology, Department of Quantum Science and Technology, The Australian National University, Canberra ACT 2601, Australia}

\author{Syed~Assad}
\affiliation{A*STAR Quantum Innovation Centre (Q.InC), Institute of Materials Research and Engineering (IMRE), Agency for Science, Technology and Research (A*STAR), 2 Fusionopolis Way, Singapore, 138634, Republic of Singapore}
\affiliation{Centre for Quantum Computation and Communication Technology, Department of Quantum Science and Technology, The Australian National University, Canberra ACT 2601, Australia}

\author{Ping~Koy~Lam}
\affiliation{A*STAR Quantum Innovation Centre (Q.InC), Institute of Materials Research and Engineering (IMRE), Agency for Science, Technology and Research (A*STAR), 2 Fusionopolis Way, Singapore, 138634, Republic of Singapore}
\affiliation{Centre for Quantum Computation and Communication Technology, Department of Quantum Science and Technology, The Australian National University, Canberra ACT 2601, Australia}

\date{\today}

\pacs{Valid PACS appear here}

\begin{abstract}
	Gaussian channel simulation is an essential paradigm in understanding the evolution of bosonic quantum states. It allows us to investigate how such states are influenced by the environment and how they transmit quantum information. This makes it an essential tool for understanding the properties of Gaussian quantum communication. Quantum teleportation provides an avenue to effectively simulate Gaussian channels such as amplifier channels, loss channels and classically additive noise channels. However, implementations of these channels, particularly quantum amplifier channels and channels capable of performing Gaussian noise suppression are limited by experimental imperfections and non-ideal entanglement resources. In this work, we overcome these difficulties using a heralded quantum teleportation scheme that is empowered by a measurement-based noiseless linear amplifier.  The noiseless linear amplification enables us to simulate a range of Gaussian channels that were previously inaccessible. In particular, we demonstrate the simulation of non-physical Gaussian channels otherwise inaccessible using conventional means. We report Gaussian noise suppression, effectively converting an imperfect quantum channel into a near-identity channel. The performance of Gaussian noise suppression is quantified by calculating the transmitted entanglement.
\end{abstract}

\maketitle

\section{Introduction}

A fundamental requirement in both quantum computation and communication is the reliable transmission of information from one point to another. Achieving this entails transferring information across a quantum channel, a complex task made more difficult by decoherence, a pervasive factor in any physical quantum system.  It was proven that a variety of practical communication channels can be accurately modelled by Gaussian channels, in which information transfer can be optimized using Gaussian encoding~\cite{wolf2007quantum}. The simulation of Gaussian channels provides an effective tool to study the evolution of an arbitrary quantum state transmitted through the channel, and is therefore of great significance for assessing the performance of a system and accordingly mitigating errors~\cite{tserkis2018simulation,laurenza2019tight,giovannetti2014ultimate,ralph2011quantum}.\\ 
Quantum teleportation offers one avenue for the exchange of quantum information, through entanglement and classical communications. In practical scenarios, teleportation can be envisioned as a quantum channel, albeit one that often falls short of achieving perfect transfer of the shared state due to inherent limitations~\cite{furusawa1998unconditional,yonezawa2004demonstration,takei2005high,yukawa2008high,takeda2013deterministic}. Specifically, in Gaussian channels, teleportation can both introduce and remove noise from a state~\cite{tserkis2018simulation,ralph2011quantum}. This property renders the teleportation protocol capable of simulating various phase-insensitive Gaussian channels, as elucidated in the work of Tserkis \textit{et al}~\cite{tserkis2018simulation}.
\begin{figure*}[t!]
  \subfloat{%
   \includegraphics[width=\linewidth]{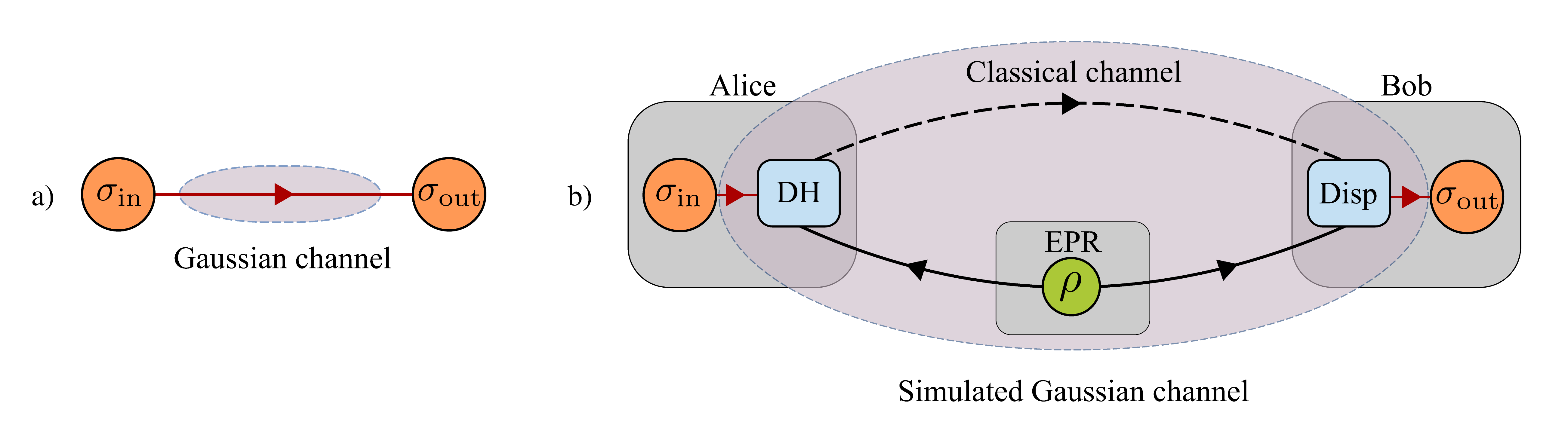}}
 \caption{Gaussian channel simulation. a) the Gaussian channel is completely characterised by the channel parameters i.e., channel transmissivity $\tau$ and noise $\nu$. b) the equivalent Gaussian channel simulated by the teleporter. DH is the dual-homodyne Alice use to measure the input state and Disp is the displacements made by Bob on the arm of the entangled state he receives. The exact experimental setup of the teleporter is shown in Fig.~\ref{experimentsetup}.}\label{simulation}
\end{figure*}
Conventional deterministic quantum teleportation has been used to simulate a wide array of channels. However, to simulate all possible bosonic Gaussian channels proposed by theories~\cite{holevo2001evaluating,giovannetti2014ultimate}, an unphysical infinite-energy resource state is required. This is the main difficulty that has thus-far prevented the simulation of arbitrary Gaussian channels.\\ 
In this work, we overcome this constraint by equipping a conventional teleporter with a probabilistic measurement-based noiseless amplifier~(MBNLA) on Alice's station. This heralded teleporter was recently demonstrated, showcasing an enhancement in teleportation fidelity as compared to conventional means~\cite{zhao2023enhancing}. Results from the previous work are constrained to identity channels, where the output resembles the input state with a finite fidelity. In this work, the teleporter operates in a much more generalised regime to enable channel simulation. This significantly broadens the operational degrees of freedom of the teleporter, rendering a variety of new phenomena possible. A noteworthy, and perhaps our most striking finding is the simulation of certain channels that were otherwise inaccessible even with infinite entanglement. We also demonstrate the transmitted entanglement can be increased without relying on additional entanglement, which allows us to correct an otherwise noisy-loss channel, i.e., Gaussian noise suppression, referred to as Gaussian error correction in Ref.~\cite{ralph2011quantum,dias2018quantum,tserkis2018simulation}. Similar entanglement distillation protocols have been demonstrated to correct lossy channels~\cite{chrzanowski2014measurement,slussarenko2022quantum}.\\
In brief, three main findings are reported in this paper. Firstly, the study simulates various known channels, including amplifier and loss channels, some of which were previously inaccessible due to limited resources for high squeezing. This was made possible by implementing heralded quantum teleportation in regimes that were unexplored in prior works~\cite{chrzanowski2014measurement,zhao2023enhancing}. Secondly, the teleportation configuration demonstrated the ability to enhance the transmitted entanglement through a loss channel, effectively implementing Gaussian noise suppression. Thirdly, owing to the probabilistic nature of the protocol, the study simulated channels that would otherwise be inaccessible, even with infinite initial entanglement.

\section{\label{level3}Gaussian channel simulation with heralded quantum teleportation}
\begin{figure*}[t!]
\centering
  \subfloat{%
   \includegraphics[width=\linewidth]{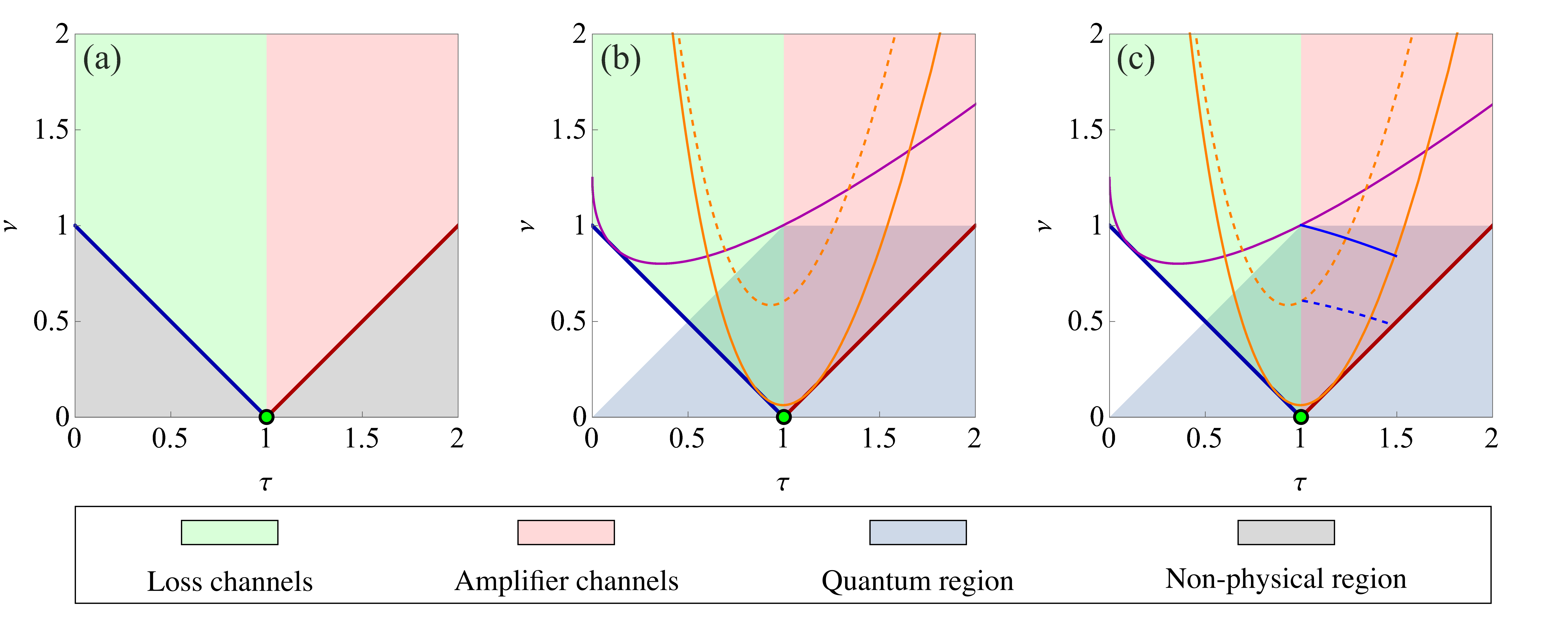}}
 \caption{Simulation of Gaussian channels. (a) The $\tau-\nu$ parameter space of Gaussian channels. The space can be divided into different regions of interest based on the values of the parameters. The green dot~($\tau=1$, $\nu=0$) corresponds to the \emph{identity channel}. Loss channels are represented by the green-shaded region and amplifier channels are represented by the red-shaded region. The red and blue solid lines correspond to \emph{pure amplifiers} and \emph{pure loss channels} respectively. The grey region represents \emph{non-physical channels}. (b)~Gaussian channels that are simulable using deterministic teleportation. The magenta line corresponds to ideal deterministic teleportation (no losses or phase noise) using 3~dB of squeezing. At this 3~dB squeezing level, the teleporter does not simulate any non-classical channels. On the other hand, the orange lines represents the channels simulated by the teleporter when using 15 dB of squeezing, with the dashed line representing the operation with 5\% loss. The quantum region, shaded in blue, is the region inaccessible to a classical teleporter~\cite{ralph1998teleportation}. A classical teleporter is essentially a measure and prepare protocol where we do not use entanglement. (c)~Gaussian channels that can be simulated using a teleporter equipped with MBNLA. The solid blue line, connecting the magenta line and the solid orange line, illustrates the range of channels simulable by a probabilistic teleporter as the MBNLA gain is progressively enhanced, while maintaining a resource EPR squeezing level of 3~dB. For a different feed-forward gain i.e., a different point on the magenta line, there are similar channels that one could simulate by tuning the MBNLA gain. In the case of a teleporter featuring 15~dB of squeezing and experiencing a 5\% loss, increasing the MBNLA gain results in the simulation of the channels represented by the dashed blue line. In both scenarios, the MBNLA gain is increased when the electronic feed-forward gain equals $\sqrt{2}$, corresponding to the unity gain regime when the MBNLA gain is set to 1.}\label{channels}
\end{figure*}
\begin{figure*}[t!]
\centering
  \subfloat{%
   \includegraphics[width=0.90\textwidth]{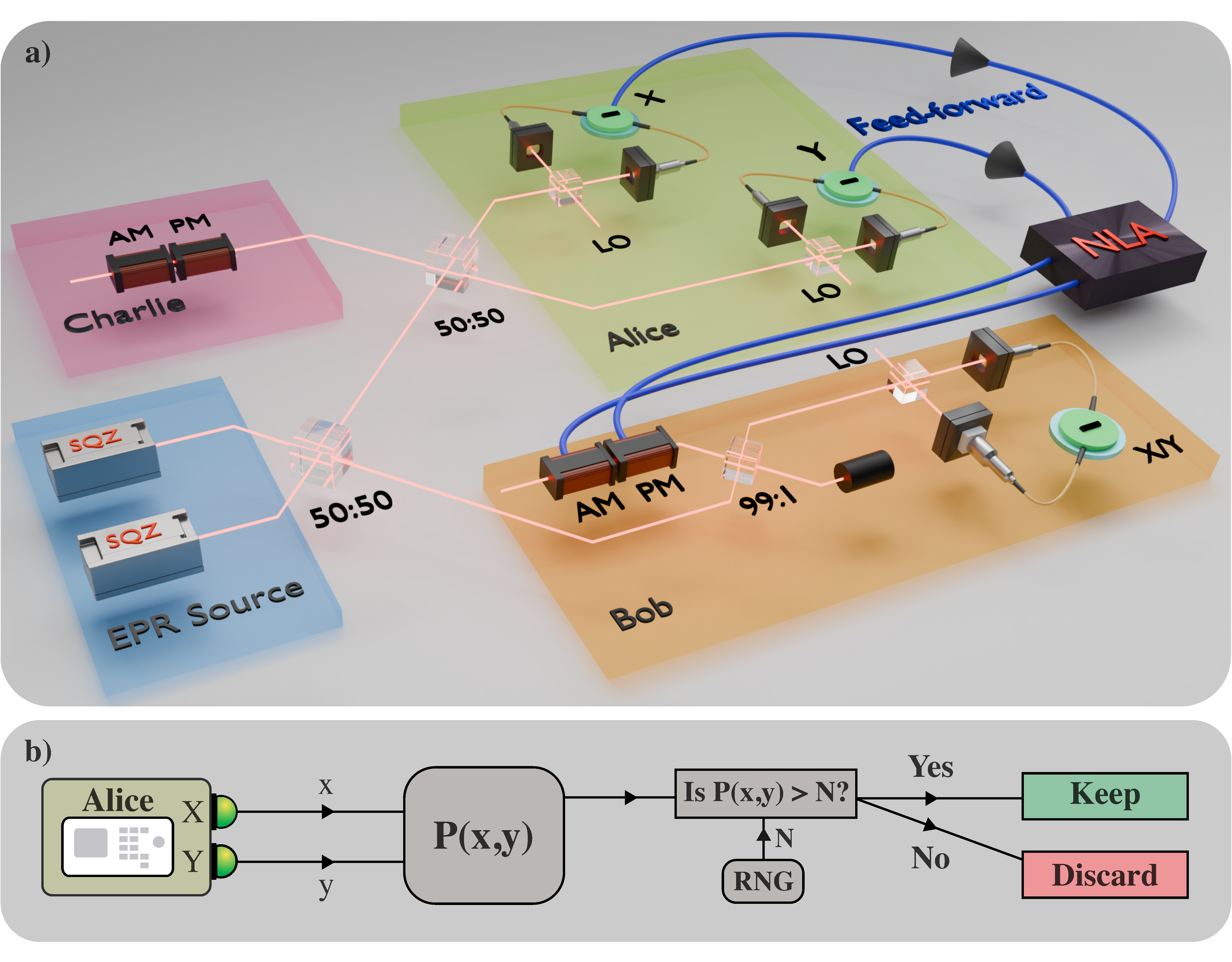}}
 \caption{Experimental scheme. (a)~Channel simulation setup based on the heralded quantum teleporter. The entanglement is created by mixing two squeezed states at a 50:50 beamsplitter. The squeezed states are generated using two optical parametric amplifiers~(SQZ)~\cite{shajilal202212}. A third party, Charlie shares a quantum state with Alice. In our experiment, we share coherent states. The coherent states are created by Charlie using a pair of amplitude~(AM) and phase electro-optic modulators~(PM). Alice performs dual homodyne measurements after mixing the state shared by Charlie with one sub-mode of the entangled state. The blue wires represents the classical channel that Alice uses to share the measurement outcomes with Bob. Bob then uses these shared classical information to reconstruct the input state sent by Charlie. Bob performs homodyne measurement to verify successful teleportation. X/Y indicates the quadrature Bob chooses to measure. LO is the local oscillator. (b)~The post-selection scheme used to implement MBNLA. The filter function P(x,y) is an inverted Gaussian as given in Eq.~\ref{filterfunction}. The random number $\mathbf{N}$ is drawn from uniform distribution and is compared with P(x,y). Based on the outcome, the measurement is kept or discarded before transmitting to Bob's station.}\label{experimentsetup} 
\end{figure*}


It is known that Gaussian channels can be faithfully simulated by quantum teleportation~\cite{ralph1998teleportation,ralph1999characterizing}, as illustrated in Fig.~\ref{simulation}. Therefore, by tuning the operating parameters of a teleporter, one could make the transformation of the state under teleportation match that of an equivalent Gaussian channel.\\
The decoherence experienced by a Gaussian state can be parameterised as a function of the channel transmissivity~$\tau$ and noise~$\nu$. The decoherence of a single mode Gaussian quantum state can be written as follows~\cite{weedbrook2012gaussian}, 
\begin{equation}
\sigma_{\text{out}} = \mathcal{U}\sigma_{\text{in}}\mathcal{U}^{\text{T}} + \mathcal{V},\label{eq1}
\end{equation}
where $\mathcal{U}=\sqrt{\tau}\mathbb{1}$ and $\mathcal{V}=\nu\mathbb{1}$ are $2\times2$ real matrices, $\mathbb{1}$ is the $2\times2$ identity matrix, and $\sigma$ is the covariance matrix. The covariance matrix is defined as defined as $\sigma_{ij} := \frac{1}{2} \langle \{ \Delta \mathbf{X}_{i}, \Delta \mathbf{X}_{j} \} \rangle$, where $\mathbf{X}$ is the amplitude quadrature operator $\hat{X}$ or phase quadrature operator $\hat{Y}$, $\Delta \mathbf{X}_{i} := \mathbf{X}_{i} - \langle \mathbf{X}_{i} \rangle$ is the standard deviation of the quadrature operator and $\{,\}$ is the anti-commutator. For amplitude quadrature,~$\hat{X} = \hat{a}^\dagger + \hat{a}$, and for phase quadrature,~$\hat{Y} = i(\hat{a}^\dagger - \hat{a}).$ The covariance matrix $\sigma$ is a real and symmetric matrix which must satisfy the uncertainty principle~\cite{weedbrook2012gaussian}. For our experiments the input state is given by a coherent state and so $\sigma = \mathbb{1}$. The restrictions on $\tau$ and $\nu$ (see below) ensure that the covariance matrix always corresponds to a physical state under the transformation given by Eq.~\ref{eq1}. The simulated Gaussian channels are completely positive and trace preserving~\cite{weedbrook2012gaussian}. Based on the values of $\tau$ and $\nu$, the Gaussian channels can be classified into the following channels as summarised in Fig.~\ref{channels}~(a): 
\begin{itemize}
\item Loss channel: The loss channel is characterised by transmissivity $\tau\in(0,1)$ and noise $\nu = (1-\tau)\chi$. For pure loss channels, $\chi=1$ and for thermal loss channels, $\chi>1$. A pure loss channel can be thought of as mixing the given Gaussian state with vacuum at a beamsplitter whereas a thermal loss channel corresponds to mixing the state with a thermal state.
\item Amplifier channel: For amplifier channels, the transmissivity $\tau$ is greater than 1. The noise added to the state is given by $\nu = (\tau-1)\chi$. For pure amplifiers, $\chi=1$ and for thermal amplifiers, $\chi>1$. Amplifier channels can be thought of as a two-mode squeezing operation with one mode being the quantum state and the other being vacuum for pure amplifier, and a thermal state for a thermal amplifier. The corresponding transmissivity is given by $\tau=\cosh{2r}$, where $r$ is the two-mode squeezing factor~\cite{caves1982quantum}.
\item Classical additive noise channel: These channels are characterised by $\tau = 1$ and $\nu>0$. This can be visualised as random displacements of the input states following a Gaussian distribution of variance $\nu$.
\item Identity channel: The identity channel is the ideal non-decohering channel where the state is perfectly shared without any added noise, i.e., $\tau=1$ and $\nu=0$. In other words, this refers to the case where the environment has not left any trace of interaction with the quantum state. 
\end{itemize}
Note that, as shown in Fig.~\ref{channels}~(a), this description of channels encompasses a non-physical region inaccessible deterministically~\cite{giovannetti2014ultimate}. In the context of this description of channels, we shall consider Gaussian noise suppression as taking an imperfect quantum channel closer to the identity channel.\\
The simulation of these Gaussian channels are performed using the heralded quantum teleporter shown in Fig.~\ref{experimentsetup}. The experiment constitutes a heralded quantum teleporter equipped with a noiseless linear amplifier, recently demonstrated in Ref.~\cite{zhao2023enhancing}. The noiseless linear amplifier~(NLA) probabilistically amplifies a coherent state $\ket{\alpha}$ to $\ket{g_{\alpha}\alpha}$, where $g_{\alpha}$ is the amplification gain~(Refer to supplementary section I for more details on NLA). We resort to a measurement-based implementation of the NLA~\cite{marek2010coherent,fiuravsek2009engineering,walk2013nondeterministic,xiang2010heralded,ferreyrol2010implementation,chrzanowski2014measurement}. The MBNLA reconstructs the probability distribution of the amplified states, therefore effectively emulates an ideal NLA. The measurement based implementation has been shown to benefit a wide range of quantum information processing tasks~\cite{zhao2023enhancing,zhao2020high,zhao2017characterization,chrzanowski2014measurement}. The implementation of the MBNLA involves the use of a filter function on the dual homodyne measurement at Alice's station as shown in Fig.~\ref{experimentsetup}. The filter function is given by~\cite{chrzanowski2014measurement,zhao2023enhancing},
\begin{equation}
P(\alpha)= \begin{cases}
    e^{\frac{1}{2}(|\alpha|^2-|\alpha_c|^2)(1-g_{\alpha}^{-2})}, & \text{if $|\alpha|<\alpha_c$}.\\
    1, & \text{otherwise}.
  \end{cases}\label{filterfunction}
\end{equation} 
where $\alpha=x+\mathbf{i}y$ is the dual homodyne outcome and $\alpha_c$ is the filter-cutoff. The gain $g_{\alpha}$ and filter-cutoff $\alpha_c$ are chosen to achieve appreciable success rates. The filter cutoff is normally between 4 to 5 standard deviations of Alice's measured state. This is to ensure that the statistics emulated by the MBNLA correspond approximately to those of an ideal NLA operation $g^{\hat{n}}$. Heralded teleportation of a state with a large mean photon number necessitates that the $\alpha_c$ is appropriately increased compared to the case where one is teleporting a vacuum state. On a successful heralding event, the amplitude and phase outcomes are rescaled by the electronic gains, $g_e$ and fed forward to the transmitted mode. Based on the rescaled dual homodyne outcomes $g_e g_{\alpha} \alpha$, Bob reconstructs the state. Refer to supplementary section II for more details on MBNLA assisted Continuous variable quantum teleportation.\\
The relation between the channel parameters~($\tau$ and $\nu$) and the teleporter operating parameters~(entanglement and feed-forward gain) is described in Ref.~\cite{tserkis2018simulation} and in the supplementary section~III. The various channels are characterised by their unique $\tau$ and $\nu$ values and by tuning the teleporter parameters, one can simulate various Gaussian channels. In the traditional teleporter, increasing the feed-forward gain~(gain-tuned operation) introduces additional noise. This additional noise hinders the simulation of certain channels, particularly quantum amplifier channels, and channels closer to the ideal identity channel. The MBNLA enables the simulation of these channels without requiring large amounts of entanglement. At sufficiently high MBNLA gain values, it is possible to simulate pure amplifier channels, a feature otherwise unattainable in the deterministic scenario. Increasing the gain further enables the  simulation of channels that is impossible even with ideal teleportation and perfect entanglement. The heralded teleporter should be also capable of increasing the transmitted entanglement through a loss channel, enabling Gaussian noise suppression. This is verified through the simulation and characterisation of various Gaussian channels shown in Fig.~\ref{channels}. Unlike existing teleportation protocols~\cite{chrzanowski2014measurement,zhao2023enhancing}, which requires the unity gain condition~($g_{\alpha}g_e =1$) for simulating identity channels~(universal teleportation), the unity gain condition is relaxed in our experiment~($g_{\alpha}g_e > 0$). More details on the non-unity gain regime of the teleporter are provided in the supplementary section~III.\\
For the simulation of quantum amplifier channels, first we perform the deterministic quantum teleportation of coherent states in the unity gain regime. Then, post-measurement, we apply the post-selection filter on the dual homodyne outcomes to perform noiseless linear amplification. Unity gain condition refers to the case where the displacement of the teleported coherent state is the same as the input coherent state. The unity gain condition is maintained during the experiment by choosing an appropriate feed-forward gain for a given entanglement used in the teleportation. The condition is independent of the displacement of the input state and therefore requires no knowledge of the input state.\\
To demonstrate Gaussian noise suppression, the loss channels were simulated using the teleporter by operating it in the gain-tuned region. In the gain-tuned region, the amplitude of the teleported coherent state is not equal to that of the input coherent state. To simulate losses, we operated with a classical feed-forward gain $g_e$ less than 1. Post-measurement, the post-selection filter is applied to perform MBNLA, which effectively implements Gaussian noise suppression.\\
We use the transmissivity $\tau$ and noise $\nu$ to characterise the simulated channels and to quantify the performance of Gaussian noise suppression, we use entanglement of formation~\cite{bennett1996mixed,tserkis2020maximum}. The amount of transmitted entanglement indicates the level of Gaussian noise suppression performed.\\
Fig~\ref{channels}~(b) plots various channels that can be simulated through deterministic teleportation at different squeezing levels. The shaded blue region signifies the unattainable region for a classical teleporter~\cite{ralph1998teleportation}. Depending on the degree of squeezing and the classical feed-forward gain, different channels can be simulated. However, for channels displaying greater quantum characteristics (meaning more amplification with no added noise or better noise suppression), it is required to have significant squeezing levels and minimal net losses. In the absence of additional losses, 3 dB of squeezing can emulate a channel at the classical limit set by the uncertainty principle~(i.e.,~$\tau>\nu~\text{and}~\nu<1$. For more details, refer to supplementary section III). However, when losses are introduced, this requires more than 3~dB of resource squeezing as the $\tau$ decreases and $\nu$ increases. When utilizing 15 dB of squeezing (represented by the solid orange line), one can simulate channels with a more pronounced quantum nature, surpassing the classical limit. Nevertheless, the dashed orange line illustrates that with a 5\% loss, the curve moves further away from the quantum realm. This is a result of the loss of correlations between the output and input states.\\
Crucially however, the incorporation of MBNLA facilitates the simulation of such channels without the stringent demand for high resource squeezing and low losses. Through the integration of MBNLA, a range of channels with higher $\tau$ and lower $\nu$ can be emulated as the MBNLA gain $g_{\alpha}$ is increased, a feat not attainable even with squeezing levels as high as 15 dB when considering realistic experimental imperfections like detection losses and imperfections in locking mechanisms. It is worth noting, however, that as the MBNLA gain increases, the probability of successful teleportation decreases. In Fig.~\ref{channels}~(c), channels simulable using the heralded quantum teleportation are represented by the dashed and solid blue lines. The solid blue line illustrates the channels simulated as the MBNLA gain is progressively increased when the teleporter operates with 3 dB of squeezing and no loss. The dashed blue line represents the channels simulated as the MBNLA gain is progressively increased when the teleporter operates with 15~dB of squeezing and 5~\% loss. The curve starts from the point at which the teleporter operates within the unity gain regime deterministically. With 3~dB of squeezing, as the MBNLA gain increases, the teleporter realises quantum channels that would otherwise be unattainable using a conventional teleporter. As the orange traces in Fig.~\ref{channels}~(b) show, the introduction of losses raises the squeezing resource requirements for simulating channels that extend beyond the classical limit, particularly in extreme cases such as pure loss channels and pure amplifiers. However, as the dashed blue line shows, the MBNLA enables us to overcome this loss and achieve a performance which cannot be achieved even in the ideal lossless case. In fact, as we show in the supplementary section~IV, MBNLA enables the simulation of virtually all channels. \\
Another way to visualise the improvement in Gaussian channel simulation is through the nature of information transfer. To help with the discussion, we introduce the \textit{joint signal transfer coefficient}~$T_q$ and the \textit{conditional variance product}~$V_q$ between the input and output states~\cite{bowen2003experimental}. The joint signal transfer coefficient is given by,
\begin{equation}
T_q = T_x + T_y = T_q = \frac{\braket{X_{\text{out}}}^2}{\braket{(\delta X_{\text{out}})^2}} + \frac{\braket{Y_{\text{out}}}^2}{\braket{(\delta Y_{\text{out}})^2}},
\end{equation}
where $T_x$ and $T_y$ are the transfer function coefficient of the respective quadratures. The transfer function coefficient is the ratio of the signal-to-noise ratio of the output to the input. $T_q$ represents the amount of information successfully recovered by Bob. The input-output conditional variance product represents the correlations between the input and output, 
\begin{equation}
V_q = V_{\text{X}_{\text{out}}|\text{X}_{\text{in}}} . V_{\text{Y}_{\text{out}}|\text{Y}_{\text{in}}},
\end{equation}
where $V_{\text{X}_{\text{out}}|\text{X}_{\text{in}}}$ and $V_{\text{Y}_{\text{out}}|\text{Y}_{\text{in}}}$ represents the input-output conditional variance of the respective quadrature. The $TV$ parameters relates to the $\tau-\nu$ parameters as follows,
\begin{equation}
V_q=\nu^2~~~\text{and}~~~T_q=\frac{2\tau}{\tau+\sqrt{V_q}}.
\end{equation}
\section{Experimental implementation and channel simulation}
\begin{figure*}[tb!]
  \subfloat{%
   \includegraphics[width=\textwidth]{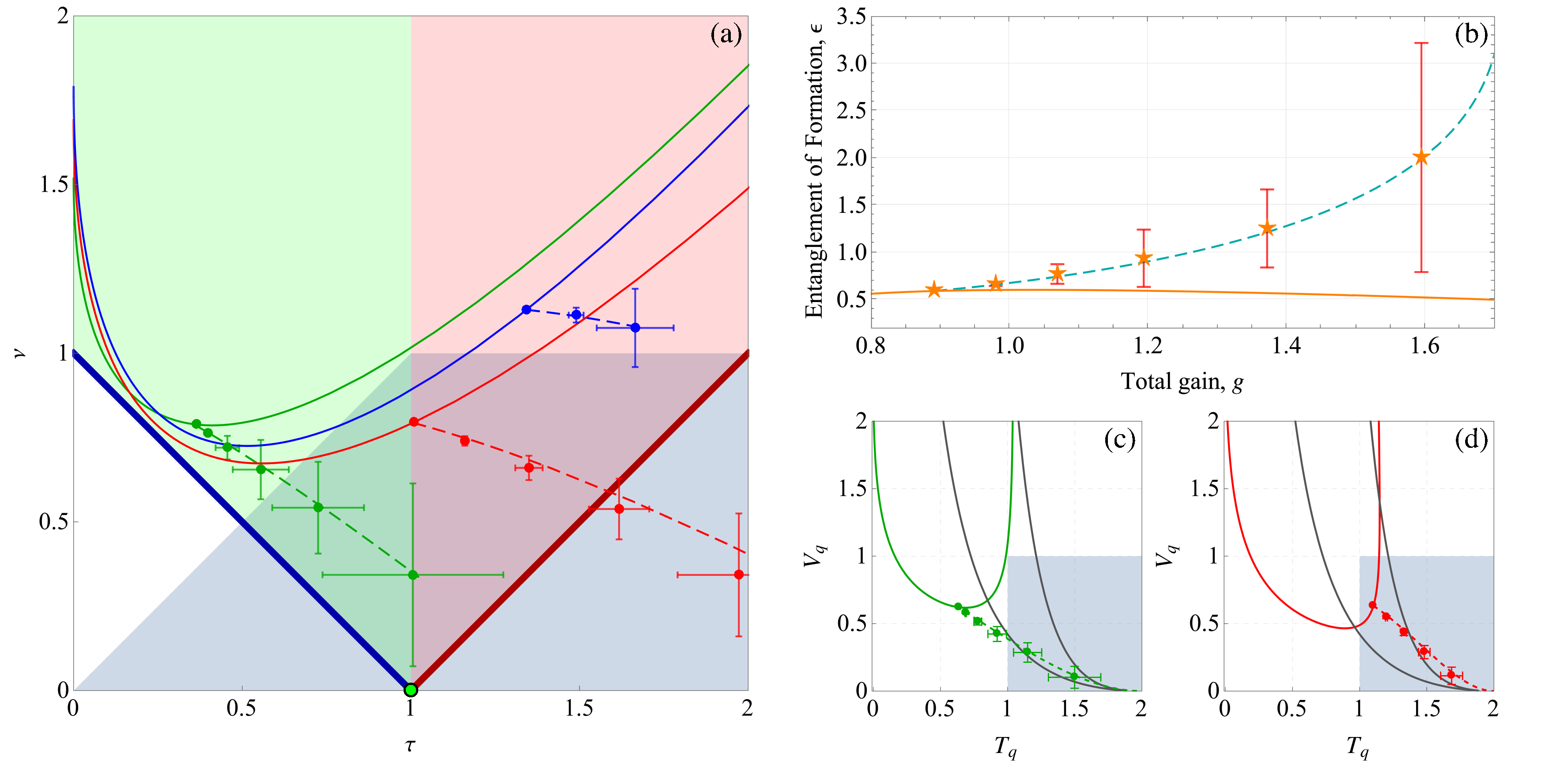}}
 \caption{Experimental results from Gaussian channel simulation using a teleporter equipped with a MBNLA. (a) depicts the characterisation results  of the simulated channels. The solid green, red, and blue curves correspond to deterministic teleportation results using 4.25 dB, 4.85 dB, and 5.15 dB squeezing, respectively. When the teleporter was operated at 5.15~dB and 4.25~dB of resource squeezing, the loss amounted to 10.5\%, whereas it was 5.5\% for 4.85~dB of squeezing. Dashed lines in the respective colours corresponds to heralded teleportation. The dots in the matching colors represent experimentally simulated channels as the MBNLA gain is increased. Simulated channels denoted by red dots and some of the green dots fall within the non-classical region of the $\tau-\nu$ diagram - channels that can only be achieved by employing higher levels of resource squeezing in a deterministic teleporter. (b) demonstrates the Gaussian noise suppression performed in the experiment. Gaussian noise suppression results in an increase in the transmitted entanglement, quantified through the entanglement of formation of the Choi state, $\epsilon$. The cyan dashed line represent $\epsilon$ as a function of $g_{\alpha}$ for a given $g_e$~(heralded teleportation) and the orange solid line is $\epsilon$ as a function of $g_e$ for $g_{\alpha}=1$~(deterministic teleportation) when the resource squeezing is 4.25~dB. The orange stars are the experimental results from Gaussian noise suppression using heralded teleportation. A substantial increase in $\epsilon$ is observed for heralded teleportation given the same resources. (c) and (d) shows the improvement in signal transfer coefficient and conditional variance product. The same colouring scheme and type of lines used in (a) are followed to represent the corresponding operating conditions and parameters. Therefore, (c) and (d) show the improvement in the TV parameters for the noise suppressing channels and amplifier channels respectively as the MBNLA gain is increased. The black lines in both (c) and (d) represent the TV parameters attained by an ideal deterministic teleporter teleporter~(no losses and phase noise) with 15~dB of resource squeezing. Trading determinism, the MBNLA equipped teleporter is capable of achieving similar information transfer characteristics with considerably less squeezing resource.}\label{results}
\end{figure*}
The experimental implementation of the Gaussian channel simulation involves the tuning of the degree of freedoms of the heralded teleporter i.e., resource squeezing, classical feed-forward gain, and the MBNLA gain. The MBNLA gain is easy to adjust and allows us to simulate a wide range of channels. Increasing the resource squeezing can also help in the process of simulating various channels. However, this causes undesirable behavior in the teleporter due to imperfections present in a realistic system, especially when combined with a high classical feed-forward gain. This undesired behavior refers to different effects that lead to a loss of correlations between the input and output. In contrast, increasing the MBNLA gain does not introduce such undesired behavior.\\
Fig.~\ref{results}~(a) illustrates the simulation of various Gaussian channels using the heralded quantum teleportation with different levels of resource squeezing. The results are benchmarked against conventional setups using deterministic teleportation and same experimental parameters.\\
We first simulate amplifier channels within the non-classical regime~(red data points). The incorporation of MBNLA heightens the non-classical characteristics of these simulated channels, going beyond the classical limits and achieving even lower $\tau$ and $\nu$ values. In this context, pure amplifiers were realized by increasing the MBNLA gain. The channel initially resembled a thermal amplifier but transformed towards a pure amplifier as the MBNLA gain increased.\\
One unique aspect of a probabilistic teleporter is its capacity to simulate channels beyond the pure amplifier channel. As shown in red in Fig.~\ref{results}~(a), we are able to simulate channels which surpass not just the classical limits and reside in the non-classical region, but remarkably enter the non-physical region described in Ref.~\cite{giovannetti2014ultimate}. These channels are only accessible probabilistically and as such do not have a deterministic physical equivalent and are not under the purview of previous studies on Gaussian channels~\cite{giovannetti2014ultimate,tserkis2018simulation,sharma2018bounding}. This is a consequence of the placement of the MBNLA in our experiment. In Ref.~\cite{tserkis2018simulation} where they consider Gaussian noise suppression~(referred to as error correction in Ref.~\cite{tserkis2018simulation,dias2018quantum,ralph2011quantum}), the maximum gain of the MBNLA is bounded based on the amount of the resource entanglement used, which is not the case in our experiment. We could implement arbitrarily large amplification surpassing such gains. However, the probabilistic nature of the protocol ensures that we do not violate any physical laws on average. Consequently, the conventional view of an amplifier channel as a two-mode squeezing operation with the other mode as a thermal state is no longer applicable. This is because the excess noise is less than vacuum noise during state amplification in the non-physical region, i.e, the noise penalty in amplifying the coherent state is less than shot-noise. Note that the same effect can be observed in channel simulation without teleportation i.e., a prepare and measure protocol with post-selection using only coherent states. However, Alice would have to share the complete information of the prepared state to Bob over a classical channel. Without the entanglement between Alice and Bob, this compromises the security of any communication protocol that is studied using such a platform and therefore we refrain from making any further comparisons between these protocols.\\
Similarly, in the case of loss channels~(green data points), MBNLA enables the simulation of channels that violated classical limits. The augmentation of MBNLA gain makes it possible to reach $\tau=1$ and $\nu=0.35$. Compared to the loss channel simulated by the deterministic teleporter, this channel ($\tau=1$,~$\nu=0.35$) represents a much closer approximation to the identity channel simulated by the teleporter equipped with the MBNLA. The underlying rationale is that as the MBNLA gain increased, the thermal background interacting with the state at the beamsplitter with transmissivity $\tau$ decreased. When $\tau=1$, the low noise in the coherent state indicates the new effective channel has only left a very minor trace of interaction with the quantum state. This can be viewed as a noise suppressing channel for Gaussian states since decoherence~(due to interacting thermal states) is circumvented by the MBNLA. One way to quantify the performance of the teleporter in performing noise suppression is to directly compare the simulated channels with an identity channel using a distance measure. However, for Gaussian channels, this is not ideal~\cite{nechita2018almost}. Instead, we look at the amount of entanglement that can be transmitted through the channel i.e., the entanglement of formation of the Choi states to quantify the performance of implemented Gaussian noise suppression. In other words, Gaussian noise suppression is the process of increasing the transmitted entanglement for the simulated channel. Fig.~\ref{results}~(b) shows the entanglement of formation~\cite{tserkis2019quantifying} of the Choi states as the MBNLA gain is increased. The dashed cyan curve in Fig.~\ref{results}~(b) shows the heralded operation of the teleporter for a given feed-forward gain, while the solid line shows the deterministic operation, i.e., the MBNLA gain is set to 1. As the MBNLA gain increases, the transmitted entanglement also increases. However, if the MBNLA gain is set to 1 and the classical gain is increased for the same total gain, the transmitted entanglement decreases resulting in an inferior Gaussian noise suppression. A similar proposal for performing Gaussian noise suppression was presented in Ref.~\cite{tserkis2018simulation}. However, in our approach, the placement of MBNLA is different. The MBNLA is placed at Alice's station instead of Bob's station. An exceptional feature of this implementation lies in how the integration of MBNLA makes it possible to perform Gaussian noise suppression exclusively utilizing physical Gaussian resources, a feat previously deemed unattainable~\cite{niset2009no} deterministically. It is also worth pointing that we do not violate any no-go theorems because of the probabilistic nature of MBNLA~\cite{eisert2002distilling,fiuravsek2002gaussian,giedke2002characterization}. We note that the use of probabilistic operations to correct lossy channels has been demonstrated previously~\cite{chrzanowski2014measurement,zhao2023enhancing,slussarenko2022quantum}. However, Ref.~\cite{zhao2023enhancing} and \cite{chrzanowski2014measurement} restricted the study to identity channels while Ref.~\cite{slussarenko2022quantum} dealt with discrete variable quantum states.\\
Finally, we analyze our experimental results in terms of information transfer, i.e., the signal transfer coefficient~($T_q$) and conditional variance product~($V_q$). The results are shown in Fig.~\ref{results}~(c) and Fig.~\ref{results}~(d). For amplifier channels, as the MBNLA gain is increased, $T_q$ increases while maintaining a strong correlation between the input and the output. The low $V_q$, beyond the classical limit, indicates the non-classical nature of the channel. The amplification is performed by the channel without tampering with the non-classical features of a state. For example, if one were to share a squeezed state using this channel, unlike a conventional amplifier that contaminates the state with at least 3 dB of noise, the quantum amplifier channel is capable of amplifying the coherent amplitude while preserving the squeezing. The effect of a loss channel is to decrease $T_q$ and increase $V_q$. The MBNLA-equipped teleporter can increase $T_q$ and decrease $V_q$, which converts a loss channel into a near-identity channel and performs Gaussian noise suppression.
\section{CONCLUSION}
We have successfully demonstrated the simulation of Gaussian channels through the utilization of continuous-variable quantum teleportation equipped with a MBNLA. Our research reveals that by introducing a MBNLA, one can reduce the resource requirements to perform useful operations like Gaussian noise suppression and the simulation of a variety of non-classical channels such as the amplifier channels. Gaussian noise suppression is executed in our experiment without any physical non-Gaussian resources. This resulted from the enhanced entanglement transmission due to the MBNLA operation. The simulated quantum amplifier channels are capable of maximising information transfer without compromising the quantum features of the shared information. Furthermore, the probabilistic nature of our protocol enables us to simulate channels which would not be accessible otherwise, even with an infinitely entangled state. These findings underscore the adaptability and promise of MBNLA-enhanced teleportation in improving the simulation of diverse Gaussian channels, each characterised by distinct noise properties.

\section*{Acknowledgements}
This research was funded by the Australian Research Council Centre of Excellence for Quantum Computation and Communication Technology (Grant No. CE110001027). P.K.L. acknowledges support from the ARC Laureate Fellowship FL150100019. This research is supported by A$^*$STAR C230917010, Emerging Technology and A$^*$STAR C230917004, Quantum Sensing.\\

\section*{Author contributions}
B.S. and L.C. conceived the project. B.S. and L.C. performed the experiment. B.S. and L.C. modelled the supporting theory and performed the numerical analysis. B.S. wrote the manuscript. All authors contributed to discussions regarding the
results in this paper. S.A and P.K.L. supervised the project.

\section*{Competing Interests}
The authors declare no competing financial or non-financial interests.

\section*{Data availability}
Data underlying the results presented in this paper are not publicly available at this time but may be obtained from the authors upon reasonable request.

\section*{Code availability}
The codes underlying the analysis presented in this paper are not publicly available at this time but may be obtained from the authors upon reasonable request.

\end{document}


\title{Supplementary Information: Improving Gaussian channel simulation using non-unity gain heralded quantum teleportation}
\maketitle
\begin{center}
\normalsize{Biveen~Shajilal$^{1,2,3}$, Lorc{\'a}n~O.~Conlon$^{1,3}$, Angus Walsh$^{3}$, Spyros~Tserkis$^{4}$,\\ Jie~Zhao$^{3}$, Jiri~Janousek$^{2,3}$, Syed~Assad$^{1,3}$, and Ping~Koy~Lam$^{1,3}$}\\
	\small{~}\\
	\small{$^{1}$\it{A*STAR Quantum Innovation Centre (Q.InC), Institute of Materials Research and Engineering (IMRE),\\ Agency for Science, Technology and Research (A*STAR), 2 Fusionopolis Way, Singapore, 138634, Republic of Singapore,}}\\
	
	\small{$^{2}$\it{Centre for Quantum Computation and Communication Technology, Research School of Engineering,\\
The Australian National University, Canberra, ACT 2601, Australia,}}\\

	\small{$^{3}$\it{Centre for Quantum Computation and Communication Technology, Department of Quantum Science and Technology,\\ The Australian National University, Canberra ACT 2601, Australia,}}\\
	\small{$^{4}$\it{Photonic Inc., Coquitlam, British Columbia, Canada.}}
	
\end{center}

\section{Noiseless Linear Amplifier}
 \begin{figure}[b!]
\center
  \includegraphics[width=0.75\textwidth]{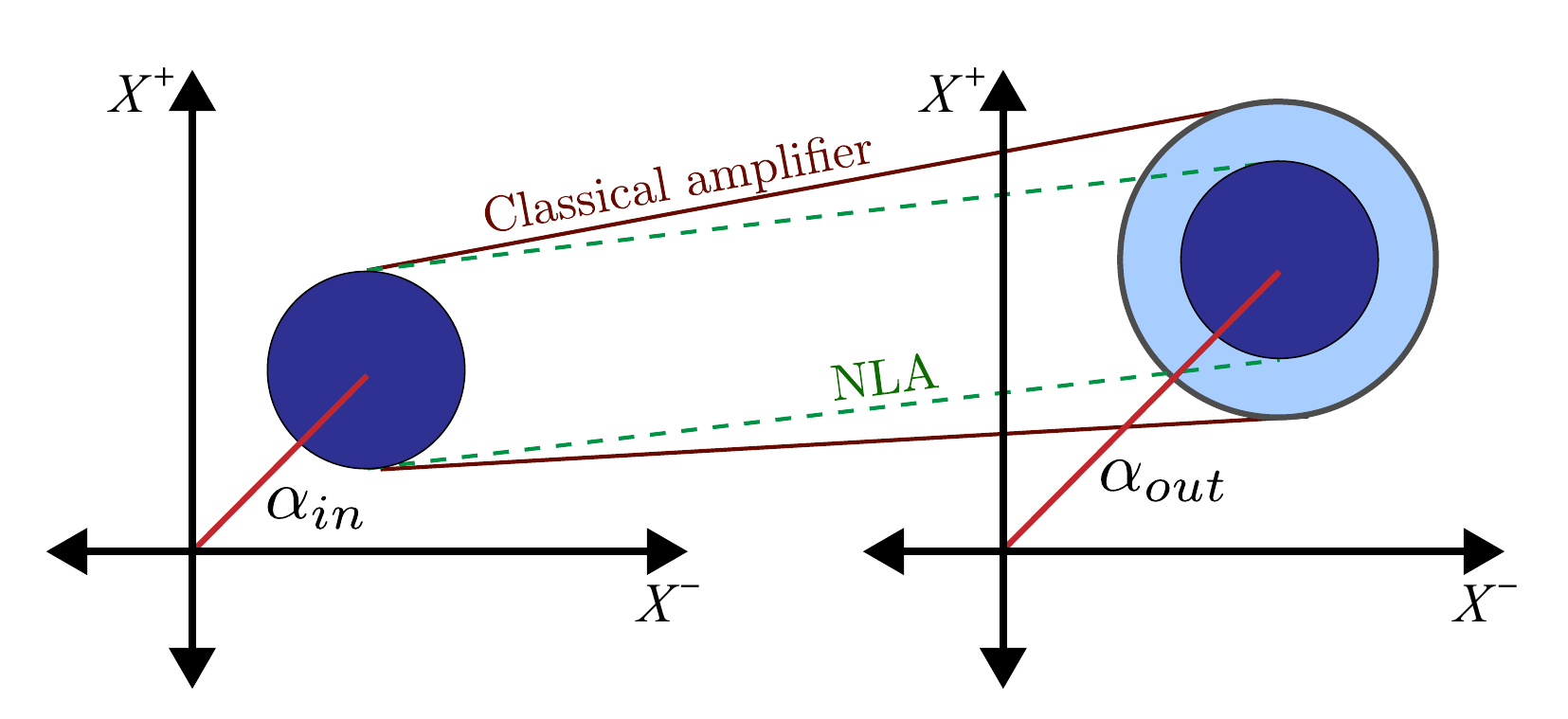}
  \caption[Amplification of coherent states]{\textbf{Amplification of coherent states.} The figure depicts the transformation of the coherent state under amplification. In the case of a classical amplifier~(shown by the red solid line), the coherent state undergoes amplification with an auxiliary mode coupled into the state resulting in a state with larger variance, i.e., a thermal state. Whereas in the case of an ideal NLA~(shown by the green dashed line), the transformation $\ket{\alpha_{in}}$ to $\ket{\alpha_{out}}$ is obtained.}
  \label{fig:NLA}
\end{figure}
Amplifiers play a crucial role in classical communications. Amplifiers amplify the signal by adding some amount of noise\footnote{$\hat{a}_{amp} = \sqrt{g}\hat{a}_{in} + \sqrt{g-1}\hat{a}_{aux}$. Often, the auxiliary mode that gets coupled into the mode under classical (or quantum-limited) amplification is the vacuum itself. Hence the noise added is in fractions of shot noise. As one could see, higher the gain $g$, the more thermal the amplified state becomes.}~\cite{caves1982quantum}. Quantum mechanically, this noise is on the order of the shot noise. The amount of shot noise added to the signal also depends on the gain of the system. Noiseless linear amplification~(NLA) refers to the amplification without incurring this noise penalty. NLA is represented by the operator $g^{\hat{\mathbf{N}}}$,
\begin{equation}
	\ket{\alpha}\rightarrow\ket{g\alpha} \equiv g^{\hat{\mathbf{N}}}\ket{\alpha}=\exp{\frac{1}{2}(g^2-1)|\alpha|^2}\ket{g\alpha}\label{NLA},
\end{equation}
where $g$ is the gain and $\hat{\mathbf{N}}$ is the number operator. The operator is an unbounded operator\footnote{An unbounded operator has a divergent set of eigenvectors. Pedantically, NLA is not even an operator for that matter.}. The  operation is also probabilistic. Otherwise, it is a direct violation of the canonical commutation results between the annihilation and creation operators. The coefficient of $\ket{g\alpha}$ in Eq.~\ref{NLA} is the probability~($P_s$) with which the state is amplified. The average state will look like, 
\begin{equation}
	\rho_{out} = P_s \ket{g\alpha}\bra{g\alpha} + (1-P_s)\ket{0}\bra{0}.\label{pp}
\end{equation}
Such a quantum limited amplifier given by Eq.~\ref{pp} corresponds to a pure amplifier~\footnote{On average, a NLA therefore does not violate the uncertainty principle or exhibit any non-physicality.}. However, by heralding on the successful events, we can benefit from the amplified states. Another unique feature of NLA is how it can be used to probabilistically increase the entanglement in an EPR state. The action of NLA results in the transformation shown below\footnote{The operation of $g^{\hat{\mathbf{N}}}$ is not represented as a unitary transformation because of the aforementioned reasons.}~\cite{chrzanowski2014measurement}, 
\begin{equation}
	g^{\hat{\mathbf{N}}}\ket{\lambda,\lambda}\bra{\lambda,\lambda}\rightarrow  P_s \ket{g\lambda,g\lambda}\bra{g\lambda,g\lambda} + (1-P_s)\ket{0,0}\bra{0,0}.
\end{equation}
In addition to being probabilistic, any practical implementation of $g^{\hat{\mathbf{N}}}$ has to be approximate as well. In all the reported schemes, a truncation of the unbounded operator is necessary.
 The better the approximation, the lower the success probability~($P_s$). Therefore, for a nominal $P_s$, high-fidelity amplification can only be performed for coherent states with very small $\bar{n}=|\alpha|^2$. The ball and stick picture depiction of the noiseless linear amplification is shown in Fig.~\ref{fig:NLA}. 

\section{MBNLA assisted Continuous variable quantum teleportation : Theoretical Model}
\begin{figure*}[t!]
\centering
  \subfloat{%
   \includegraphics[width=0.9\textwidth]{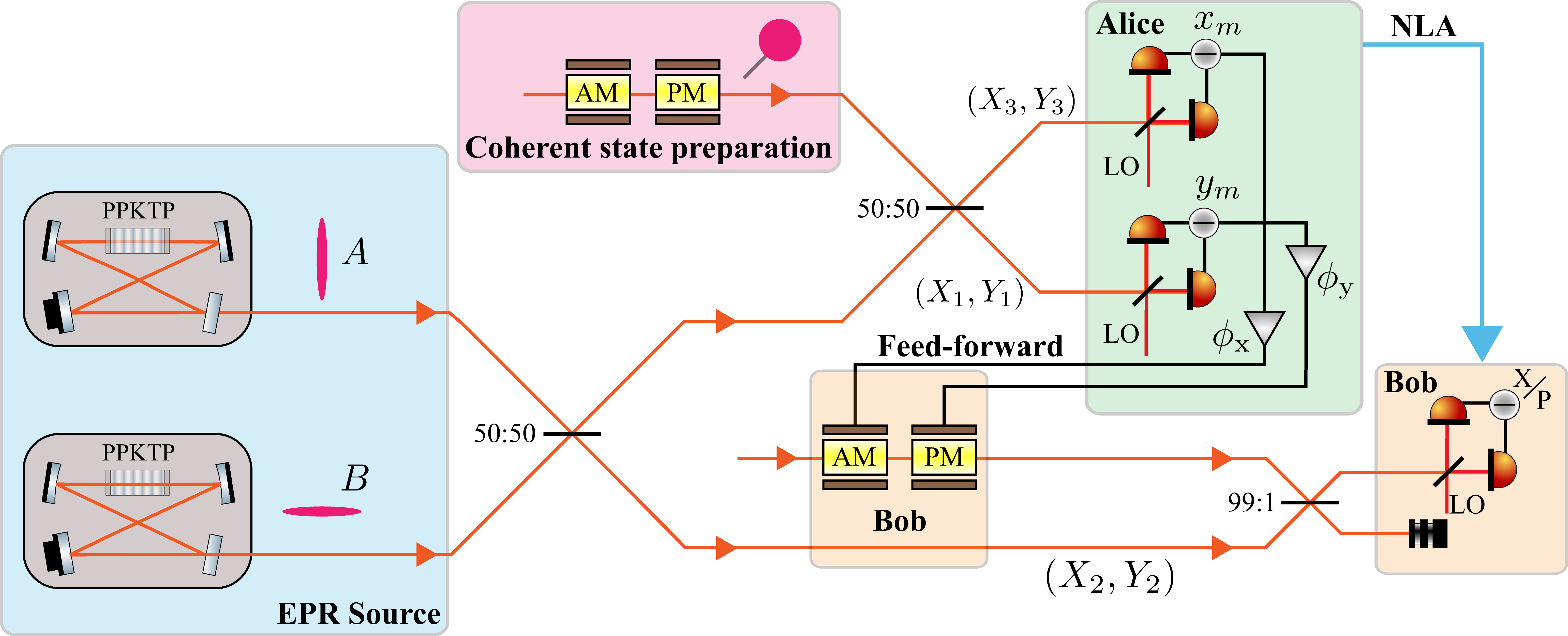}}
 \caption{\textbf{Schematic of the heralded continuous variable quantum teleporter.} $X_{n}$ and $Y_{n}$ indicates the amplitude and phase quadratures of mode $n$. The input state $\rho_{in}$ is prepared by Charlie and mixed with one arm of the EPR state resulting in mode 1 and 3. Alice performs dual homodyne measurement on these modes. The measured outcomes are shared to Bob via the classical channel. The measured variables are rescaled by the classical electronic gains $\phi_{\text{x}}$ and $\phi_{\text{y}}$. Bob uses these rescaled outcomes to displace the other arm of the EPR mode i.e., mode 2 and performs homodyne measurement on the displaced mode to verify teleportation.}\label{experimentsetup}
 \end{figure*}

Figure.~\ref{experimentsetup}(a) depicts the teleportation setup used in the experiment. A third party prepares the quadrature entanglement and shares it between Alice and Bob. Alice mixes the coherent state to be teleported with her arm of the entangled state at a 50:50 beamsplitter~(BS) and then performs a dual homodyne measurement. The measurement outcomes are then shared with Bob through a classical communication channel. Based on the measurement outcome Bob receives, he performs local displacement on the auxiliary beam and mixes it with the other mode of the EPR state at a 99:1 BS to reconstruct the input coherent state. This is equivalent to performing the displacement on the EPR mode that Bob received. Since we are teleporting Gaussian states, the mean and variance of the output state characterises the output state. The theoretical mode follows the covariance method used in section I of the supplementary information of Ref.~\cite{zhao2023enhancing}.\\ 

We start by calculating the covariance matrix of the EPR state. The covariance matrix of the EPR state is given by, 
\begin{align}
 \label{eq:HQT1C9}
 \mathbf{C} = 
 \left( \begin{array}{cccc}
 C_{11}  & 0 & C_{13} &0 \\ 
  0 &  C_{22} & 0 & C_{24} \\ 
    C_{13} & 0 & C_{11} & 0 \\
    0 &   C_{24} & 0 & C_{22}  \\
 \end{array} \right). \
\end{align}

The elements of the covariance matrix are given by, 
\begin{align}
    \label{eq:HQT2C9}
        C_{11} & = \frac{ e^{-2 r_{\mathrm{Ax}} } + e^{ 2r_{\mathrm{B x}} } }{2} , \quad C_{22} =  \frac{ e^{ -2r_{\mathrm{B y}} }  + e^{2r_{\mathrm{A y}}}  }{2},  \\[5pt]
        C_{13} & = \frac{ e^{2r_{\mathrm{B x}}} - e^{-2 r_{ \mathrm{Ax}}}  }{2} , \quad C_{24} = \frac{ e^{ -2r_{\mathrm{B y}} } - e^{2r_{\mathrm{A y}}}  }{2}   .\
    \end{align}

\noindent The $r_{\mathrm{Ax(y)}}$ and $r_{\mathrm{Bx(y)}}$ refer to the squeezing parameters of squeezed states $A$ and $B$, respectively, while the subscript denotes the respective quadrature. The input state with quadrature variances $\langle (\Delta X_{\mathrm{in}} (Y_{\mathrm{in}}))^2 \rangle$ is then mixed with one mode of the two-mode EPR state using a 50:50 BS. The combined covariance matrix is given by, 
\begin{align}
 \label{eq:HQT3C9}
 \mathbf{C'} {=} 
 \left( \begin{array}{cccccc}
 C_{11}  & 0 & C_{13}/\sqrt{2} & 0 &  -C_{13}/\sqrt{2}  &  0  \\ 
  0 &  C_{22} & 0 & C_{24}/ \sqrt{2}  & 0 &  -C_{24}/ \sqrt{2}   \\ 
    C_{13}/ \sqrt{2} & 0 &  ( \langle (\Delta X_{\mathrm{in}})^2 \rangle +  C_{11}) / 2 & 0 & (\langle (\Delta X_{\mathrm{in}})^2 \rangle - C_{11} )/2  & 0   \\
    0 &   C_{24} / \sqrt{2} & 0 & ( C_{22} + \langle ( \Delta Y_{\mathrm{in}})^2 \rangle )/2 & 0 & (\langle (\Delta Y_{\mathrm{in}})^2 \rangle- C_{22})/2  \\
    -C_{13}/\sqrt{2} & 0 & (\langle (\Delta X_{\mathrm{in}})^2 \rangle - C_{11})/2 & 0 & (C_{11} + \langle ( \Delta X_{\mathrm{in}})^2 \rangle)/2 & 0 \\
    0 & -C_{24}/\sqrt{2} & 0 & (\langle (\Delta Y_{\mathrm{in}})^2 \rangle- C_{22})/2 & 0 & (C_{22} + \langle (\Delta Y_{\mathrm{in}})^2 \rangle)/2 \\
 \end{array} \right) . \
\end{align}
Elements of the covariance matrix $\mathbf{C'}$ in Eq.~\eqref{eq:HQT3C9} are given by $$\mathbf{C'}_{ij} := \frac{1}{2} \langle \{ \Delta \mathbf{X}_{i}, \Delta \mathbf{X}_{j} \} \rangle,$$ where $\Delta \mathbf{X}_{i} := \mathbf{X}_{i} - \langle \mathbf{X}_{i} \rangle$ and $\{,\}$ is the anti-commutator. If one wants to look at the variance of a particular mode, say mode 1, this is given by $C_{11}{=}\frac{1}{2} \langle \{ \Delta X_{1}, \Delta X_{1} \} \rangle $. Similarly correlations exhibited between modes can be analysed by looking at the covariance term. Covariance between mode $n$ and $m$ is given by $\langle \Delta X_{n} \Delta X_{m} \rangle$. For ease of representation, the use of $\Delta$ before the quadrature variables will be omitted in the following discussions and equations. After the input state\footnote{In our experiment we teleport coherent states.} is mixed with the EPR mode at the 50:50 BS, the outgoing modes 1 and 3 are subjected to dual homodyne measurement i.e., the amplitude and phase quadratures are measured simultaneously. The other submode of the EPR state shared with Bob evolves based on Alice's measurement outcome $\alpha_{m} = (x_{m} + i y_{m})/\sqrt{2}$. The conditional mean and variance of the amplitude quadrature of the transmitted EPR mode are given by,
\begin{equation}
 \label{eq:HQT4C9}
    \begin{aligned}
       \mu_{X_{2} | \bar{x}_{m}} & = \langle X_{2} \rangle +\frac{ \langle X_{2} X_{3} \rangle }{ \langle X_{3} X_{3} \rangle } \left( \bar{x}_{m} - \langle X_{3} \rangle \right) =  \frac{ -\sqrt{2} C_{13}  }{ \langle X_{\mathrm{in}} X_{\mathrm{in}} \rangle   + C_{11}  } \left( \bar{x}_{m} -\frac{\langle X_{\mathrm{in}} \rangle  }{\sqrt{2}} \right)^{T}   ,   \\[5pt]
       \Sigma_{X_{2} | \bar{x}_{m}} & = \langle X_{2} X_{2} \rangle - \frac{\langle X_{2} X_{3} \rangle \langle X_{3} X_{2} \rangle  }{ \langle X_{3}X_{3} \rangle  } =  C_{11} - \frac{ C_{13}^2 }{ \langle X_{\mathrm{in}} X_{\mathrm{in}} \rangle +  C_{11}  } . \
    \end{aligned}
\end{equation}
Similarly, the conditional mean and variance of the  phase quadrature of the transmitted EPR mode are given by, 
\begin{equation}
 \label{eq:HQT5C9}
    \begin{aligned}
       \mu_{Y_{2} | \bar{y}_{m}} & = \langle Y_{2} \rangle +\frac{ \langle Y_{2} Y_{1} \rangle }{ \langle Y_{1} Y_{1} \rangle } \left( \bar{y}_{m} - \langle Y_{1} \rangle \right) =  \frac{ \sqrt{2} C_{24}  }{ \langle Y_{\mathrm{in}} Y_{\mathrm{in}} \rangle + C_{22}  } \left( \bar{y}_{m} -\frac{\langle Y_{\mathrm{in}} \rangle  }{\sqrt{2}} \right)^{T}   ,   \\[5pt]
        \Sigma_{Y_{2} | \bar{y}_{m}} & = \langle Y_{2} Y_{2} \rangle - \frac{\langle Y_{2} Y_{1} \rangle \langle Y_{1} Y_{2} \rangle  }{ \langle Y_{1}Y_{1} \rangle  } =  C_{22} - \frac{ C_{24}^2 }{ \langle Y_{\mathrm{in}} Y_{\mathrm{in}} \rangle +C_{22}  } .  \
    \end{aligned}
\end{equation}
Integrating the conditional mean and variance over all possible outcomes $x_m$ gives the average mean of the transmitted EPR mode. Conditional variance also indicates how strong the entanglement is between the two EPR modes. The conditional variance of one mode of an EPR state given a measurement outcome on the other mode is given by, 
\begin{equation}
 \label{eq:CondV}
    \begin{aligned}
      \Sigma_{X_{\mathrm{a}}|\bar{x}_{\mathrm{b}}} = \langle X_{\mathrm{a}} X_{\mathrm{a}}\rangle \left(  1- \frac{\langle X_{\mathrm{a}} X_{\mathrm{b}}\rangle^2  }{ \langle X_{\mathrm{a}} X_{\mathrm{a}} \rangle \langle X_{\mathrm{b}} X_{\mathrm{b}} \rangle   }\right) = \frac{1}{\text{cosh}(2r)},\
    \end{aligned}
\end{equation}
 where $r$ is the squeezing parameter used in the generation of the EPR state. This indicates that when $r$ asymptotes to infinity, Alice can estimate the amplitudes with zero variance, given the knowledge of Bob's measurement outcome $\bar{x}_{\mathrm{b}}$. In the case of our teleporter, where we are interested in teleporting coherent states (or classical mixtures of coherent states), when Alice performs the dual homodyne measurement, the transmitted mode is projected onto a coherent state (assuming the squeezers are creating same levels of pure squeezing).\\
 
\begin{figure*}[h!]
\centering
  \subfloat{%
   \includegraphics[width=\textwidth]{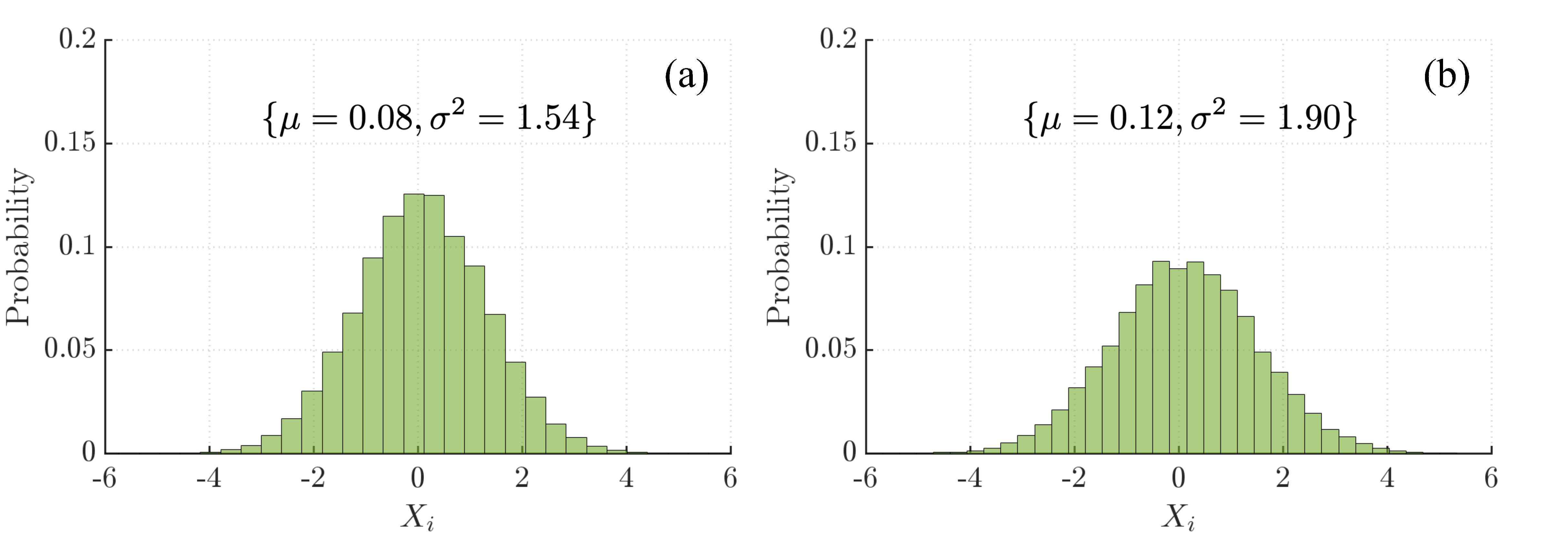}}
 \caption{\textbf{Quadrature measurement histograms of Bob's homodyne measurement.} The teleporter is operated to simulate amplifier channels in the non-classical region~(Refer main text for more details on channels and non-classical region). (a) shows the histogram of homodyne measurements performed by Bob before post selection. (b) shows the histogram after the post-selection filter is applied. Increasing the post-selection gain increases the amplification resulting in a distribution with a larger variance. Increasing the post-selection gain results in a smaller dataset. The probability of success~(given by Eq.~\ref{ff1}) was $0.0016$. At higher gains, because of the low probability of success, the dataset size becomes considerably smaller resulting in larger error bars for the figure of merits used used in this study.}\label{bob}
 \end{figure*}

 The implementation of NLA at Alice's measurement station probabilistically increases the entanglement distributed between Alice and Bob. The implementation of the physical-NLA however introduces other complexities\footnote{One suggested method of implementing NLA involves the use of quantum scissors~\cite{ralph2009nondeterministic,tserkis2018simulation}.}. Here we instead use a measurement based NLA~(MBNLA). The MBNLA albeit not being a physical emulation of the physical NLA, can reconstruct the measurement statistics of an ideal NLA. The measurement based implementation has previously demonstrated numerous experimental advantages~\cite{zhao2020high,chrzanowski2014measurement,haw2016surpassing}. The implementation of MBNLA involves the use of a filter function on the heterodyne measurement at Alice's station. The acceptance probability of the filter function is given by,
\begin{equation}
f(\alpha_m)= \begin{cases}
    e^{\frac{1}{2}~(|\alpha_m|^2-|\alpha_c|^2)(1-g^{-2})}, & \text{if $\alpha_m<\alpha_c$}.\\
    1, & \text{otherwise}.
  \end{cases}\label{filterfunction}
\end{equation}
where $\alpha=x+\mathbf{i}y$ is the heterodyne outcome and $\alpha_c$ is the filter-cutoff. The gain $g$ and filter-cutoff $\alpha_c$ are chosen to achieve improvement in the figure of merits being considered while maintaining appreciable post-selection rates. The filter-cutoff is normally between $4-5$ standard deviations of Alice's measured state. This is to ensure that the statistics emulated by the MBNLA correspond to a physical-NLA $g^{\hat{n}}$ that approximates well for given state with the same mean photon number. Heralded teleportation of a state with larger mean photon number would mean that the $\alpha_c$ should be appropriately increased compared to the case where one is teleporting vacuum\footnote{One does not need to know the exact encoding of the input state. Knowledge of the alphabet being encoded is sufficient to set the appropriate $\alpha_c$.}. On a successful heralding event, the amplitude and phase outcomes are rescaled by the electronic gains~($g_e = \phi_{\text{x(y)}}$) and fed forward to the transmitted mode. Based on the rescaled heterodyne outcomes, Bob reconstructs the state\footnote{The homodyne outcomes are rescaled by the total gain, which is the product of electronic gain and the MBNLA gain, i.e., $gg_e$.}. The process of applying the filter function in the experiment is as follows, 
\begin{enumerate}
	\item Alice performs the dual homodyne measurement i.e., Alice measure $x_m$ and $y_m$. 
	\item Alice calculates the value of $f(\alpha_m)$ based on the measured $x_m$ and $y_m$.
	\item The calculated $f(\alpha_m)$ is compared with a randomly drawn number $N$ between 0 and 1.  If the value of $f(\alpha_m)$ is greater than $N$, the outcome is retained and fed-forward to Bob after rescaling by an electronic gain~$\phi_{x(y)}$.
\end{enumerate}
The probability of getting a measurement outcome $\alpha_m$,  is proportional to $e^{-|\alpha_{m} - \alpha_{0}|^2}$ and the output distribution after post-selection can therefore be written down as, 
\begin{equation}
	p(\alpha_m)f(\alpha_m) = \begin{cases}
    Me^{-|\alpha_{m} - \alpha_{0}|^2}e^{\frac{1}{2}~(|\alpha_m|^2-|\alpha_c|^2)(1-g^{-2})}, & \text{if $\alpha_m<\alpha_c$}.\\
    Me^{-|\alpha_{m} - \alpha_{0}|^2}, & \text{otherwise}.
  \end{cases},\label{ff1}
\end{equation}
where $M$ is the normalisation factor. The post selection therefore emulates the effect of NLA i.e., amplify the mean and variance of quadrature distribution by $g^2$ and are given by, 
\begin{align}
 \label{eq:HQT6C9}
 \langle \tilde{X}_{3} \rangle = g^2 \langle X_{3} \rangle, \quad \langle  \tilde{X}_{3} \tilde{X}_{3} \rangle = g^2  \langle  X_{3} X_{3} \rangle = g^2 (\langle  X_{\mathrm{in}} X_{\mathrm{in}} \rangle + C_{11})/ 2,\\ 
 \langle \tilde{Y}_{1} \rangle = g^2 \langle Y_{1} \rangle, \quad \langle \tilde{Y}_{1} \tilde{Y}_{1} \rangle = g^2  \langle  Y_{1} Y_{1} \rangle = g^2 ( \langle  Y_{\mathrm{in}} Y_{\mathrm{in}}  \rangle  + C_{22})/2,\ 
\end{align} 
where $\tilde{X}_{1}$ and $\tilde{Y}_{1}$ are the quadrature operators after post-selection. The post-selection has no effect on the conditional statistics exhibited by the EPR mode. Therefore we can write\footnote{The same holds for the phase quadrature.}, 
\begin{align}
 \label{eq:HQT8C9}
\frac{ \langle \tilde{X}_{2} \tilde{X}_{3}\rangle }{ \langle \tilde{X}_{3} \tilde{X}_{3} \rangle  } =  \frac{ \langle X_{2} X_{3} \rangle}{ \langle X_{3}X_{3} \rangle  },
	\quad \langle \tilde{X}_{2} \tilde{X}_{2} \rangle - \frac{\langle \tilde{X}_{2} \tilde{X}_{3} \rangle \langle \tilde{X}_{3} \tilde{X}_{2} \rangle}{\langle \tilde{X}_{3} \tilde{X}_{3} \rangle} = \langle X_{2}X_{2} \rangle - \frac{ \langle X_{2} X_{3} \rangle \langle X_{3} X_{2} \rangle}{ \langle X_{3} X_{3} \rangle }.
\end{align}
Using Eq.~\ref{eq:HQT8C9}, \ref{eq:HQT6C9} and \ref{eq:HQT3C9}, we can write, 
 \begin{equation}
 \label{eq:HQT9C9a}
 	\tilde{\mu}_{X_{2} | \bar{\tilde{x}}_{m}} = \mu_{X_{2}| \bar{x}_{m}}  +  \frac{ \langle \tilde{X}_{2} \tilde{X}_{3} \rangle }{ \langle \tilde{X}_{3} \tilde{X}_{3} \rangle } \left( g^2 \bar{x}_{m} - \frac{ \langle X_{\mathrm{in}} \rangle }{\sqrt{2}}  \right)  =\frac{ C_{13} \left( 1- g^2  \right) \langle X_{\mathrm{in}} \rangle }{ \langle X_{\mathrm{in}} X_{\mathrm{in}} \rangle  + C_{11} }.
 \end{equation}
Similarly, 
\begin{equation}
\label{eq:HQT9C9b}
	\tilde{\mu}_{Y_{2}| \bar{\tilde{y}}_{m} } = \mu_{Y_{2} | \bar{y}_{m} } + \frac{ \langle \tilde{Y}_{2} \tilde{Y}_{1}\rangle }{ \langle \tilde{Y}_{1} \tilde{Y}_{1} \rangle } ( g^2 \bar{y}_{m} - \frac{ \langle Y_{\mathrm{in}} \rangle }{ \sqrt{2} }  ) = \frac{ C_{24} \left( g^2 -1  \right) \langle Y_{\mathrm{in}} \rangle }{  \langle Y_{\mathrm{in}} Y_{\mathrm{in}} \rangle +  C_{22} }.
\end{equation} 	
 After post-selection, the signals are rescaled by the electronic gains $\phi_{x(y)}$ and Bob performs displacement operation\footnote{Amplitude and phase modulation using a pair of electrooptic modulators.} on his EPR mode. The mean and variance of the teleported state can be written as, 
 \begin{equation}
 	\label{eq:HQT10C9a}
  \langle X_{\mathrm{out}} \rangle =  \tilde{\mu}_{X_{2}| \bar{\tilde{x}}_{m}} + \langle \tilde{X}_{3}\rangle = 
  \tilde{\mu}_{X_{2}| \bar{\tilde{x}}_{m}} + \phi_{x} g^2 \langle X_{\mathrm{in}} \rangle/ \sqrt{2},~~\langle X_{\mathrm{out}}^2 \rangle = \langle \tilde{X}_{2} \tilde{X}_{2} \rangle + 2\phi_{x} \langle \tilde{X}_{2} \tilde{X}_{3}\rangle + \phi_{x}^2 g^2 ( \langle X_{\mathrm{in}} X_{\mathrm{in}} \rangle + C_{11})/2 
 \end{equation}
 and 
 \begin{equation}
 	\label{eq:HQT10C9b}
  \langle Y_{\mathrm{out}} \rangle = \tilde{\mu}_{Y_{2}| \bar{\tilde{y}}_{m}} + \langle \tilde{Y}_{1}\rangle
  = \tilde{\mu}_{Y_{2}| \bar{\tilde{y}}_{m}} + \phi_{y} g^2 \langle Y_{\mathrm{in}} \rangle / \sqrt{2} ,~~\langle Y_{\mathrm{out}}^2 \rangle = \langle \tilde{Y}_{2} \tilde{Y}_{2} \rangle + 2\phi_{y} \langle \tilde{Y}_{2} \tilde{Y}_{1} \rangle + \phi_{y}^2 g^2 ( \langle Y_{\mathrm{in}} Y_{\mathrm{in}} \rangle + C_{22})/2 .\end{equation}\\
  The process of applying the post selection filter was mentioned before. Once the post-selection filter is applied on Alice's data, based on the post-selection, the quadrature distribution of Bob's data changes as given by Eq.~\ref{eq:HQT10C9a} and \ref{eq:HQT10C9b}. The histogram of the Bob's data before Alice applies the MBNLA filter and Bob's data after post-selection is shown in Fig.~\ref{bob}.\\
  
  The amount of data for a given experimental run goes down with higher MBNLA gain, according to the probability of success. The probability of success corresponding to the data showed in Fig.~\ref{bob}(b) was $0.0016$. This is one of the main limitations of any probabilistic protocol. One needs to figure out the right experimental parameters where they observe a sufficient benefit from post-selection before the probability of success reaches unfeasible values. The probability of success is given by multiplying the distribution of $\alpha_{m}$ with the filter function from Eq.~\ref{ff1},
  \begin{equation}
    	P_{\mathrm{s}} = \frac{e^{(g-1)|\alpha_{m}|^2}}{e^{\alpha_{\mathrm{c}}^2(1-1/g)}} \iint\limits_{\left|\alpha\right|< \alpha_{\mathrm{c}}} \exp\left( \frac{-\left|\alpha- g \alpha_{m}
  \right|^2}{g} \right)  \text{d}^2\alpha +  \iint\limits_{\left|\alpha \right|\geq\alpha_{\mathrm{c}}} \exp\left(-\left|
      \alpha- \alpha_{m} \right|^2 \right) \text{d}^2\alpha.
  \end{equation}
\section{Characterising Teleportation in the non-unity gain regime: Transfer function coefficient, Conditional variance and channel parameters}
The efficacy of teleportation is often quantified using the overlap between the input and output states i.e., fidelity. For Gaussian input states, the fidelity is given by,
\begin{multline}
\label{eq:FUG1C9}
 \mathcal{F} = \frac{2}{\sqrt{( \langle  X_{\mathrm{out}} X_{\mathrm{out}} \rangle  + \langle X_{\mathrm{in}} X_{\mathrm{in}} \rangle)    ( \langle  Y_{\mathrm{out}} Y_{\mathrm{out}} \rangle + \langle Y_{\mathrm{in}} Y_{\mathrm{in}} \rangle )} } \\ \text{exp} \left[ -\frac{1}{2} \left(  \frac{ (\langle X_{\mathrm{out}} \rangle - \langle X_{\mathrm{in}} \rangle  )^2  }{       ( \langle  X_{\mathrm{out}} X_{\mathrm{out}} \rangle  + \langle X_{\mathrm{in}} X_{\mathrm{in}} \rangle) }  +    \frac{ (\langle Y_{\mathrm{out}} \rangle - \langle Y_{\mathrm{in}} \rangle  )^2  }{      ( \langle  Y_{\mathrm{out}} Y_{\mathrm{out}} \rangle + \langle Y_{\mathrm{in}} Y_{\mathrm{in}} \rangle )   }   \right)   \right] .\
\end{multline}
As one can see, the fidelity is dependent on the input amplitude when $\langle X_{\mathrm{out}} \rangle \neq \langle X_{\mathrm{in}} \rangle $ and $\langle Y_{\mathrm{out}} \rangle \neq \langle Y_{\mathrm{in}} \rangle $, which is the case when the teleporter is operated beyond the unity gain regime~($\langle X_{\mathrm{out}} \rangle = \langle X_{\mathrm{in}} \rangle $ and $\langle Y_{\mathrm{out}} \rangle = \langle Y_{\mathrm{in}} \rangle $). In this work, we are interested in operating and characterising the teleporter beyond the unity gain regime in the interest of simulating various Gaussian channels and fidelity is therefore not an ideal figure of merit\footnote{Other drawbacks of using Fidelity are covered in detail by Tserkis et al.~\cite{tserkis2018simulation}.}. We can benchmark the performance of our teleporter from a quantum communication perspective using the \textit{joint signal transfer coefficient}~$T$ and the \textit{conditional variance product}~$V$ between the input and output states. Other than being state-independent, the $TV$ measure comes with the added benefit of being invariant under local operations and also reveals the exact information transfer coefficient during teleportation. This is crucial in the case where the output contains all the necessary information for an accurate reconstruction while the fidelity still indicates the process to be inefficient. Another instance where such a figure is deemed useful is where with no prior knowledge, one could characterise how efficient the teleportation process is in retaining the non-classical features of the input states. Analogous to how the performance of quantum non-demolition measurement is quantified, the teleportation can be characterised through the signal transfer coefficient~(through $T$) and the non-classical correlation between the input and output~(through $V$). The joint signal transfer coefficient is given by,
\begin{equation}
T_q = T_x + T_y = \frac{\text{SNR}_{\text{xout}}}{\text{SNR}_{\text{xin}}} + \frac{\text{SNR}_{\text{yout}}}{\text{SNR}_{\text{yin}}},
\end{equation}
where $T_x$ and $T_y$ are the transfer function coefficient of the respective quadratures. $T_q$ represents the amount of information successfully recovered by Bob. $T_q$ can be written down in terms of mean and variance of the quadrature operators as follows,
\begin{equation}
T_q = \frac{\braket{X_{\text{out}}}^2}{\braket{ X_{\text{out}}^2}}\frac{\braket{ X_{\text{in}}^2}}{\braket{X_{\text{in}}}^2} + \frac{\braket{Y_{\text{out}}}^2}{\braket{ Y_{\text{out}}^2}}\frac{\braket{ Y_{\text{in}}^2}}{\braket{Y_{\text{in}}}^2}
\label{eq1}.
\end{equation}
The input-output conditional variance product represents the correlations between the input and output, 
\begin{equation}
V_q = V_{\text{X}_{\text{out}}|\text{X}_{\text{in}}} . V_{\text{Y}_{\text{out}}|\text{Y}_{\text{in}}}\label{eq2}
\end{equation}
where $V_{\text{X}_{\text{out}}|\text{X}_{\text{in}}}$ and $V_{\text{Y}_{\text{out}}|\text{Y}_{\text{in}}}$ represents the input-output conditional variance of the respective quadrature, which we previously calculated for an ideal heralded teleporter. The $TV$ parameters for classical teleportation are bounded by $T_q<1$ and $V_q>1$. This is due to the noise penalty imposed by Heisenberg's uncertainty principle under simultaneous measurement of the conjugate variables. Therefore one could think of the criteria for successful teleportation in terms of $TV$ as $T$ being greater than 1 and $V$ less than 1, implying the respective measures surpassed their respective classical limits. The perfect reconstruction of the state requires complete transfer of the unknown input state to be received by Bob and the output state is indentical to the input state. This corresponds to $T_q = 2$ and $V_q = 0$. This is achieved for an ideal teleporter with perfect EPR correlations that operates with infinitely squeezed states. In brief, the quantumness of the teleporter is beyond attaining good fidelity, especially beyond the unity-gain regime and the state-independent measure helps one to reveal this. In case of a teleporter with no loss, we can use Eq.~\ref{eq1},~\ref{eq2}, \ref{eq:HQT10C9a} and \ref{eq:HQT10C9b} to calculate $T_q$ and $V_q$ as a function of $\langle X_{\text{in}}\rangle$, $\phi_{x(y)}$, $g$, and $r$\footnote{The squeezing parameter of the two optical parametric amplifiers generating the EPR state and the electronic feedforward gain of the teleporter was precisely chosen to ensure 
 $T_x = T_y$ and $V_{\text{X}_{\text{out}}|\text{X}_{\text{in}}} = V_{\text{Y}_{\text{out}}|\text{Y}_{\text{in}}}$.},
\begin{equation}
T_q = \frac{\langle X_{\text{in}}\rangle^2 \left(\sqrt{2}~g^2 \phi_x -2 \left(g^2-1\right) \tanh (r)\right)^2}{g^2 \left(\left(\phi_x^2+2\right) \cosh (2 r)-2 \sqrt{2} \phi_x   \sinh (2 r)\right)+g^2 \left(\phi_x ^2-2\right)+2}
\label{eq3},
\end{equation}
and
\begin{dmath}
V_q = \frac{1}{16} \left[ \langle X_{\text{in}}\rangle^2 \left(\sqrt{2} g^2 \phi_x  -2 \left(g^2-1\right) \tanh (r)\right)^2-4 g^2 \left(\phi_x   \cosh (r)-\sqrt{2} \sinh (r)\right)^2+8 \sinh ^2(r)-4 \cosh (2 r)\right]^2
\label{eq4}.
\end{dmath}
So far the results presented represents an ideal heralded teleporter, in practice, there are experimental imperfections that affects the teleportation protocol. The optical propagation loss in Bob's EPR mode and Bob's homodyne detection efficiency were major sources of losses and a crucial experimental imperfection. This is modelled by coupling a vacuum mode into Bob's EPR mode through a beamsplitter transformation~\cite{weedbrook2012gaussian}. The measured total efficiency of our system was T = 0.89~\footnote{For ease of representation, propagation loss plus homodyne efficiency is represented as one term.}. Using these losses and by following the same covariance matrix approach, we can re-derive Eq.~\ref{eq3} and \ref{eq4}. The new $T_q$ and $V_q$ are given by,
\begin{dmath}
T_q = \frac{2 \left(\frac{g^2 \sqrt{T} \langle X_{\text{in}}\rangle \left(\frac{\left(e^{-2 r}-e^{2 r}\right) \sqrt{T}}{2 \sqrt{2} \left(\frac{1}{2} \sqrt{T} \cosh (2 r)-\frac{\sqrt{T}}{2}+1\right)}+\phi_x \right)}{\sqrt{2}}-\frac{\left(e^{-2 r}-e^{2 r}\right) T \langle X_{\text{in}}\rangle}{4 \left(\frac{1}{2} \sqrt{T} \cosh (2 r)-\frac{\sqrt{T}}{2}+1\right)}\right)^2}{g^2 \left(\frac{1}{2} \sqrt{T} \cosh (2 r)-\frac{\sqrt{T}}{2}+1\right) \left(\frac{\left(e^{-2 r}-e^{2 r}\right) \sqrt{T}}{2 \sqrt{2} \left(\frac{1}{2} \sqrt{T} \cosh (2 r)-\frac{\sqrt{T}}{2}+1\right)}+\phi_x \right)^2-\frac{\left(e^{-2 r}-e^{2 r}\right)^2 T}{8 \left(\frac{1}{2} \sqrt{T} \cosh (2 r)-\frac{\sqrt{T}}{2}+1\right)}+\frac{1}{2} \left(e^{-2 r}+e^{2 r}\right)}
\label{eq5},
\end{dmath}
and
\begin{dmath}
V_q = \frac{1}{16} \left(\frac{T \langle X_{\text{in}}\rangle^2 \left(\sqrt{T} \left(2 \left(g^2-1\right) \sinh (2 r)-\sqrt{2} g^2 \phi_x  \cosh (2 r)\right)+\sqrt{2} g^2 \left(\sqrt{T}-2\right) \phi_x \right)^2}{\left(\sqrt{T} \cosh (2 r)-\sqrt{T}+2\right)^2}-\frac{2 g^2 \left(\sqrt{T} \left(\sqrt{2} \sinh (2 r)-\phi_x  \cosh (2 r)\right)+\left(\sqrt{T}-2\right) \phi_x \right)^2}{\sqrt{T} \cosh (2 r)-\sqrt{T}+2}+\frac{4 T \sinh ^2(2 r)}{\sqrt{T} \cosh (2 r)-\sqrt{T}+2}-4 \cosh (2 r)\right)^2
\label{eq6}.
\end{dmath}
\begin{figure}[t!]
\centering
  \subfloat{
   \includegraphics[width=0.8\linewidth]{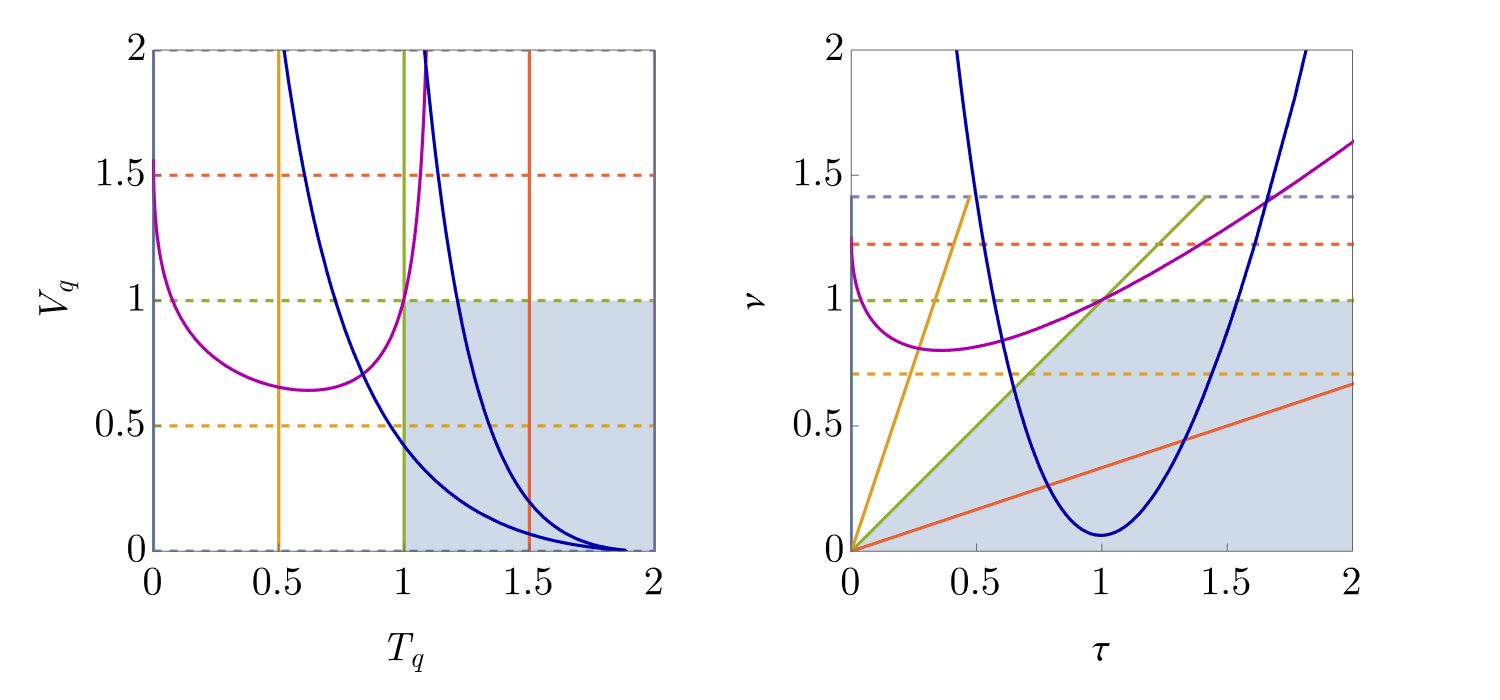}}
 \caption[The mapping between the $TV$ and $\tau\nu$ parameter space]{\textbf{The mapping between the $TV$ and $\tau\nu$ parameter space}. This map indicates that measuring the transfer function coefficient along with correlations between the input and input gives us information on the equivalent channel parameters. The shaded region corresponds to the quantum region that exceeds the classical limit of the respective measures. The blue and magenta curves represent the curves of the channels simulated for an ideal deterministic teleporter operating with 15~dB and 3~dB squeezing respectively. The classical feedforward gain is varied to simulate the various channels. By changing the resource squeezing level and electronic gain, one could simulate other channels of interest.}\label{map}
\end{figure}
Characterising teleportation in the context of simulating Gaussian channels would be much more useful if one could represent the transformation of the state induced by the teleporter in terms of the physical quantities such as transmissivity $\tau$ and noise $\nu$, see Eq.~(1) and section II (A) in the main text for more details on channel parameters. In fact, for a coherent state input, $TV$ parameters can be mapped to $\tau\nu$ parameters and one could intuitively understand the transformation of a quantum state induced by said channel. The map between the two parameter spaces is given by,
\begin{equation}
\nu=\sqrt{V_q}~~~\text{and}~~~\tau=\frac{T_q\sqrt{V_q}}{2-T_q}.\label{eq7}
\end{equation}\\
From Eq.~\ref{eq5},~\ref{eq6} and ~\ref{eq7}, it is clear that for a given coherent state input with amplitude quadrature mean $\langle X_{\text{in}}\rangle$, one could control the channel parameters by tuning parameters $\phi_{x(y)}$, $g$, and $r$. Fig.~\ref{map} shows the set of channels a deterministic teleporter~(MBNLA gain, g=1) can simulate by varying the electronic gain for different levels of squeezing. Since both the parameter space represents the same information regarding the nature of the teleporter~(and the quantumness), we rely primarily on the $\tau\nu$ measure as it is the simplest way to visualise the corresponding channels.

\section{Measurement based noiseless linear amplifier and Channel simulation}
\begin{figure}[t!]
\centering
  \subfloat{
   \includegraphics[width=0.7\linewidth]{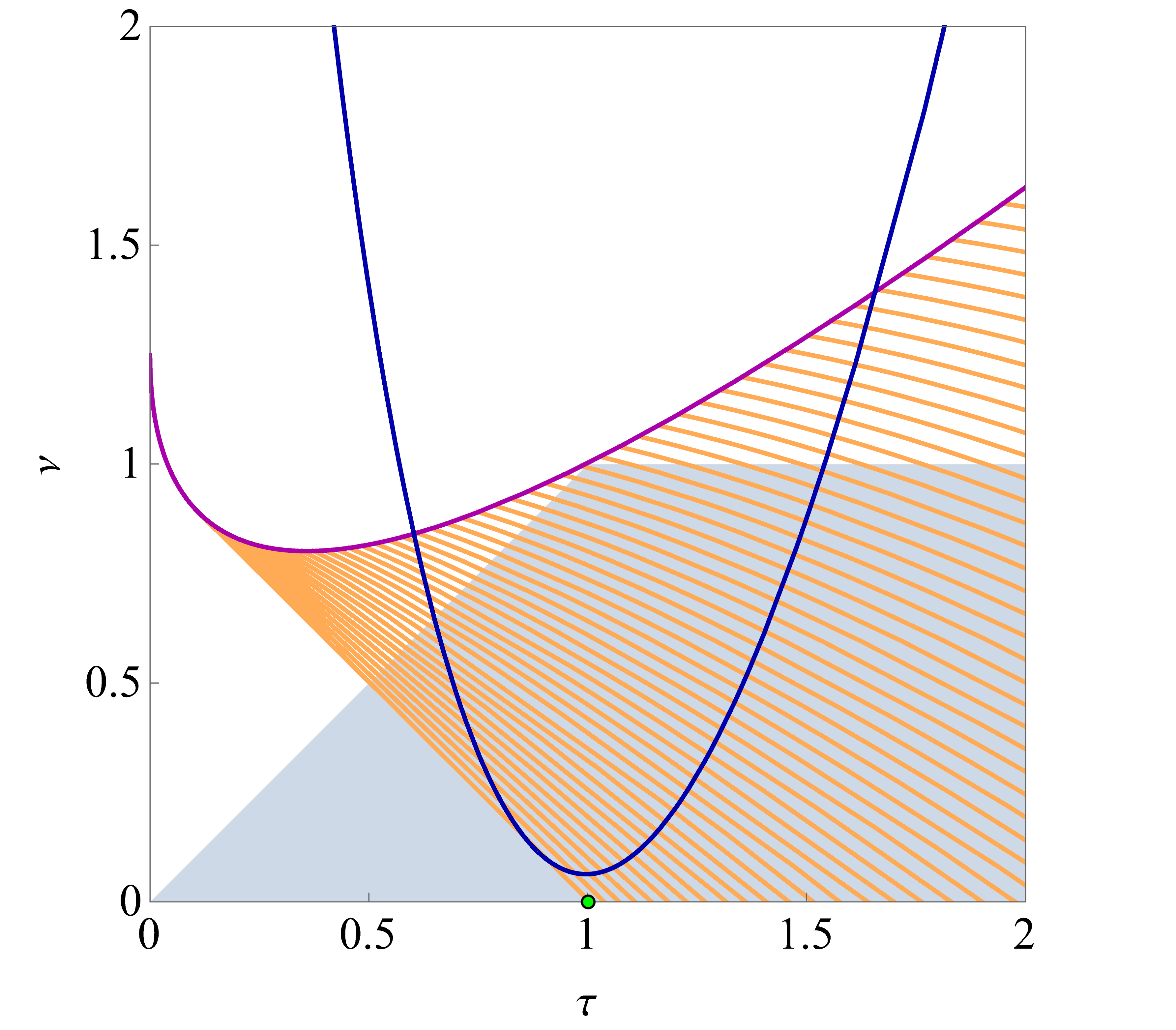}}
 \caption{\textbf{Comparison between deterministic and heralded teleportation equipped with MBNLA}. The blue and magenta curves represent the set of the channels simulated by an ideal deterministic teleporter operating with 15~dB and 3~dB squeezing respectively. The electronic gain is varied to simulate the various channels on that line. The green dot represents an identity channel and the blue shaded region corresponds to the quantum region that exceeds the classical limit of the respective measures. Each orange line represents a set of channels simulated by an ideal heralded teleporter for a given electronic gain and with 3~dB of squeezing. Unlike a deterministic teleporter, a heralded teleporter can simulate various channels by just controlling the post-selection gain and electronic gain with minimal squeezing resources.}\label{mbnla}
\end{figure}

In Fig.~\ref{map}, for a given squeezing level and coherent state input, different channels are simulated by varying $\phi_{x(y)}$. Channels that exhibit non-classical behaviour are easily simulated when the squeezing levels are high. However, by equipping the deterministic teleporter with MBNLA, we can simulate such channels without requiring high levels of squeezing. Moreover, the MBNLA gain acts as an added degree of freedom in controlling the teleporter for the purpose of channel simulation. In the deterministic scenario, one needs to tune both $r$ and $\phi_{\text{x(y)}}$ to simulate different regions of the channel whereas in the case of the MBNLA, channels with highly non-classical behaviour can be simulated by just changing the combination of classical gain and MBNLA gain. In addition to the fact that one can simulate channels with higher non-classical characteristics with lower squeezing, in a practical setting, it is ideal to change these two parameters over the squeezing parameter to simulate different channels of interest. For instance, increasing the resource squeezing can cause undesirable behaviour in the teleporter due to the imperfections\footnote{Particularly, squeezing impurity, locking instabilities, phase noises and losses~\cite{shajilal202212}.} present in a realistic system, especially when combined with a high classical feed-forward gain. This undesired behaviour leads to different effects leading to a loss of the correlations between the input and output. Fortunately, increasing the MBNLA gain does not introduce such undesired behaviour. In our approach for simulating various channels, we take advantage of this resilience by operating the deterministic teleporter with low resource squeezing. Fig.~\ref{mbnla} shows the comparison of channels that are simulated by an ideal deterministic teleporter operating with 15~dB and 3~dB of squeezing and ideal heralded teleporter\footnote{Ideal heralded teleporter refers to an heralded teleporter with zero losses. The case of zero losses is shown to indicate that we are not just circumventing experimental limitations through the implementation of MBNLA.} equipped with MBNLA, operating with 3~dB of squeezing. The magenta line represents the channels simulable by an ideal teleporter operating with 3 dB of squeezing. Each point on the magenta line represents a channel simulated for a certain value of the electronic gain. By equipping the teleporter with MBNLA, for the same value of electronic gain, one could simulate another set of channels indicated by the orange lines originating from these points\footnote{For ease of representation, the lines were drawn for finite steps of electronic gain. In practise, the MBNLA is capable of simulating any channels in the region of the orange lines from the figure.} by increasing the MBNLA gain. With 3~dB of squeezing, the heralded teleporter is capable of simulating channels ranging from error correcting channels~(the set of points on the orange lines that satisfies the condition $\tau_{\text{initial}}<\tau_{\text{final}}$ and  $\nu_{\text{initial}}>\nu_{\text{final}}$\footnote{$\tau_{\text{initial}}$ indicates the $\tau$ before post-selection and $\tau_{\text{final}}$ indicates the $\tau$ after post-selection. The same notation is used for $\nu$ as well.}) to amplifier channels~($\tau>1$ and $\nu<1$). It is worth noting that a deterministic teleporter operating with even 15~dB of squeezing is incapable of simulating a identity channel indicated by the green dot at $\tau=1$ and $\nu=0$, whereas an heralded teleporter is capable of simulating a channel very close to an identity channel, even with only 3~dB of squeezing.\\

It is worth noting that all of these improvements come at the cost of determinism. As the gain is increased, the number of successful events decreases, or in other words, the probability of success is reduced. As a consequence, the channel parameters are calculated from a smaller sample size each time as the gain is increased. This results in large error bars in our results for very high gain values. However, this can be compensated for by leaving the experiment running longer with the same amount of resource entanglement, which would result in a larger sample size. Another way to compensate for this issue is to increase the resource entanglement\footnote{In other words, a higher degree of squeezing should be used in the generation of EPR states.}. This would mean that you do not need the same amount of high gain to simulate the same set of channels. This would naturally result in a higher probability of success and smaller error bars. Ideally, repeating the experiment with a high duty cycle squeezer that is capable of generating very high levels of squeezing would increase the overall quality of the results, i.e., decrease the size of the error bars. MBNLA still remains useful for such an experiment because it allows us to simulate certain channels that are inaccessible even with perfect initial EPR correlations.